\documentclass[journal,draftclsnofoot,onecolumn,12pt]{IEEEtran}

\usepackage{graphicx}
\usepackage{subfigure}
\usepackage{multirow}
\usepackage{array}
\newcolumntype{P}[1]{>{\centering\arraybackslash}p{#1}}
\newcolumntype{M}[1]{>{\centering\arraybackslash}m{#1}}

\usepackage{amsthm,amssymb,amsmath,graphicx,multirow,color,amsfonts}%
\usepackage[update,prepend]{epstopdf}
\usepackage{multirow}
\usepackage[latin1]{inputenc}
\usepackage{tikz}
\usepackage{bbm} % for \mathbbm{1}
\usepackage{pdfpages}
\usepackage{multirow}
\usepackage{subfig}
\usepackage{comment}
 \usepackage{graphicx}
\usepackage{multicol}
\usepackage{cite}

\usepackage[justification=centering]{caption}
\usepackage{textcomp}
\usepackage{psfrag}
\usepackage{arydshln}
\usepackage{url}
\usepackage{soul}
\usepackage{graphicx,color}
\usepackage[nolist]{acronym}
\usepackage{algorithm,algorithmic} %algorithm
\usepackage{mathtools,lipsum}
\usepackage{cuted}
\usepackage{amsmath}
\newtheorem{proposition}{Proposition} 
\usepackage{mathrsfs}

\usepackage[capitalise]{cleveref}
\Crefname{equation}{Eq.\!}{Eqs.\!}
\Crefname{figure}{Fig.\!}{Figs.\!}
\Crefname{tabular}{Tab.\!}{Tabs.\!}
\Crefname{section}{Section\!}{Sections.\!}

\def\nb0{{\mathbf{0}}}
\def\nb1{{\mathbf{1}}}

\newtheorem{definition}{Definition}

\def\argmin{\operatorname{arg~min}}
\def\argmax{\operatorname{arg~max}}
\newenvironment{sequation}{
\begin{equation}\small}{\end{equation}
}

\begin{document}
% \receiveddate{XX Month, XXXX}
% \reviseddate{XX Month, XXXX}
% \accepteddate{XX Month, XXXX}
% \publisheddate{XX Month, XXXX}
% \currentdate{XX Month, XXXX}
% \doiinfo{XXXX.2023.XXXXXXX}
%\pagenumbering{gobble}
\graphicspath{{./Figures/}}
	\begin{acronym}

\acro{5G-NR}{5G New Radio}
\acro{3GPP}{3rd Generation Partnership Project}
\acro{ABS}{aerial base station}
\acro{AC}{address coding}
\acro{ACF}{autocorrelation function}
\acro{ACR}{autocorrelation receiver}
\acro{ADC}{analog-to-digital converter}
\acrodef{aic}[AIC]{Analog-to-Information Converter}     
\acro{AIC}[AIC]{Akaike information criterion}
\acro{aric}[ARIC]{asymmetric restricted isometry constant}
\acro{arip}[ARIP]{asymmetric restricted isometry property}

\acro{ARQ}{Automatic Repeat Request}
\acro{AUB}{asymptotic union bound}
\acrodef{awgn}[AWGN]{Additive White Gaussian Noise}     
\acro{AWGN}{additive white Gaussian noise}

\acro{APSK}[PSK]{asymmetric PSK} 

\acro{waric}[AWRICs]{asymmetric weak restricted isometry constants}
\acro{warip}[AWRIP]{asymmetric weak restricted isometry property}
\acro{BCH}{Bose, Chaudhuri, and Hocquenghem}        
\acro{BCHC}[BCHSC]{BCH based source coding}
\acro{BEP}{bit error probability}
\acro{BFC}{block fading channel}
\acro{BG}[BG]{Bernoulli-Gaussian}
\acro{BGG}{Bernoulli-Generalized Gaussian}
\acro{BPAM}{binary pulse amplitude modulation}
\acro{BPDN}{Basis Pursuit Denoising}
\acro{BPPM}{binary pulse position modulation}
\acro{BPSK}{Binary Phase Shift Keying}
\acro{BPZF}{bandpass zonal filter}
\acro{BSC}{binary symmetric channels}              
\acro{BU}[BU]{Bernoulli-uniform}
\acro{BER}{bit error rate}
\acro{BS}{base station}
\acro{BW}{BandWidth}
\acro{BLLL}{ binary log-linear learning }

\acro{CP}{Cyclic Prefix}
\acrodef{cdf}[CDF]{cumulative distribution function}   
\acro{CDF}{Cumulative Distribution Function}
\acrodef{c.d.f.}[CDF]{cumulative distribution function}
\acro{CCDF}{complementary cumulative distribution function}
\acrodef{ccdf}[CCDF]{complementary CDF}               
\acrodef{c.c.d.f.}[CCDF]{complementary cumulative distribution function}
\acro{CD}{cooperative diversity}

\acro{CDMA}{Code Division Multiple Access}
\acro{ch.f.}{characteristic function}
\acro{CIR}{channel impulse response}
\acro{cosamp}[CoSaMP]{compressive sampling matching pursuit}
\acro{CR}{cognitive radio}
\acro{cs}[CS]{compressed sensing}                   
\acrodef{cscapital}[CS]{Compressed sensing} %will not include it in the list
\acrodef{CS}[CS]{compressed sensing}
\acro{CSI}{channel state information}
\acro{CCSDS}{consultative committee for space data systems}
\acro{CC}{convolutional coding}
\acro{Covid19}[COVID-19]{Coronavirus disease}

\acro{DAA}{detect and avoid}
\acro{DAB}{digital audio broadcasting}
\acro{DCT}{discrete cosine transform}
\acro{dft}[DFT]{discrete Fourier transform}
\acro{DR}{distortion-rate}
\acro{DS}{direct sequence}
\acro{DS-SS}{direct-sequence spread-spectrum}
\acro{DTR}{differential transmitted-reference}
\acro{DVB-H}{digital video broadcasting\,--\,handheld}
\acro{DVB-T}{digital video broadcasting\,--\,terrestrial}
\acro{DL}{DownLink}
\acro{DSSS}{Direct Sequence Spread Spectrum}
\acro{DFT-s-OFDM}{Discrete Fourier Transform-spread-Orthogonal Frequency Division Multiplexing}
\acro{DAS}{Distributed Antenna System}
\acro{DNA}{DeoxyriboNucleic Acid}

\acro{EC}{European Commission}
\acro{EED}[EED]{exact eigenvalues distribution}
\acro{EIRP}{Equivalent Isotropically Radiated Power}
\acro{ELP}{equivalent low-pass}
\acro{eMBB}{Enhanced Mobile Broadband}
\acro{EMF}{ElectroMagnetic Field}
\acro{EU}{European union}
\acro{EI}{Exposure Index}
\acro{eICIC}{enhanced Inter-Cell Interference Coordination}

\acro{FC}[FC]{fusion center}
\acro{FCC}{Federal Communications Commission}
\acro{FEC}{forward error correction}
\acro{FFT}{fast Fourier transform}
\acro{FH}{frequency-hopping}
\acro{FH-SS}{frequency-hopping spread-spectrum}
\acrodef{FS}{Frame synchronization}
\acro{FSsmall}[FS]{frame synchronization}  
\acro{FDMA}{Frequency Division Multiple Access}

\acro{GA}{Gaussian approximation}
\acro{GF}{Galois field }
\acro{GG}{Generalized-Gaussian}
\acro{GIC}[GIC]{generalized information criterion}
\acro{GLRT}{generalized likelihood ratio test}
\acro{GPS}{Global Positioning System}
\acro{GMSK}{Gaussian Minimum Shift Keying}
\acro{GSMA}{Global System for Mobile communications Association}
\acro{GS}{ground station}
\acro{GMG}{ Grid-connected MicroGeneration}

\acro{HAP}{high altitude platform}
\acro{HetNet}{Heterogeneous network}

\acro{IDR}{information distortion-rate}
\acro{IFFT}{inverse fast Fourier transform}
\acro{iht}[IHT]{iterative hard thresholding}
\acro{i.i.d.}{independent, identically distributed}
\acro{IoT}{Internet of Things}                      
\acro{IR}{impulse radio}
\acro{lric}[LRIC]{lower restricted isometry constant}
\acro{lrict}[LRICt]{lower restricted isometry constant threshold}
\acro{ISI}{intersymbol interference}
\acro{ITU}{International Telecommunication Union}
\acro{ICNIRP}{International Commission on Non-Ionizing Radiation Protection}
\acro{IEEE}{Institute of Electrical and Electronics Engineers}
\acro{ICES}{IEEE international committee on electromagnetic safety}
\acro{IEC}{International Electrotechnical Commission}
\acro{IARC}{International Agency on Research on Cancer}
\acro{IS-95}{Interim Standard 95}

\acro{KPI}{Key Performance Indicator}

\acro{LEO}{low earth orbit}
\acro{LF}{likelihood function}
\acro{LLF}{log-likelihood function}
\acro{LLR}{log-likelihood ratio}
\acro{LLRT}{log-likelihood ratio test}
\acro{LoS}{Line-of-Sight}
\acro{LRT}{likelihood ratio test}
\acro{wlric}[LWRIC]{lower weak restricted isometry constant}
\acro{wlrict}[LWRICt]{LWRIC threshold}
\acro{LPWAN}{Low Power Wide Area Network}
\acro{LoRaWAN}{Low power long Range Wide Area Network}
\acro{NLoS}{Non-Line-of-Sight}
\acro{LiFi}[Li-Fi]{light-fidelity}
 \acro{LED}{light emitting diode}
 \acro{LABS}{LoS transmission with each ABS}
 \acro{NLABS}{NLoS transmission with each ABS}

\acro{MB}{multiband}
\acro{MC}{macro cell}
\acro{MDS}{mixed distributed source}
\acro{MF}{matched filter}
\acro{m.g.f.}{moment generating function}
\acro{MI}{mutual information}
\acro{MIMO}{Multiple-Input Multiple-Output}
\acro{MISO}{multiple-input single-output}
\acrodef{maxs}[MJSO]{maximum joint support cardinality}                       
\acro{ML}[ML]{maximum likelihood}
\acro{MMSE}{minimum mean-square error}
\acro{MMV}{multiple measurement vectors}
\acrodef{MOS}{model order selection}
\acro{M-PSK}[${M}$-PSK]{$M$-ary phase shift keying}                       
\acro{M-APSK}[${M}$-PSK]{$M$-ary asymmetric PSK} 
\acro{MP}{ multi-period}
\acro{MINLP}{mixed integer non-linear programming}

\acro{M-QAM}[$M$-QAM]{$M$-ary quadrature amplitude modulation}
\acro{MRC}{maximal ratio combiner}                  
\acro{maxs}[MSO]{maximum sparsity order}                                      
\acro{M2M}{Machine-to-Machine}                                                
\acro{MUI}{multi-user interference}
\acro{mMTC}{massive Machine Type Communications}      
\acro{mm-Wave}{millimeter-wave}
\acro{MP}{mobile phone}
\acro{MPE}{maximum permissible exposure}
\acro{MAC}{media access control}
\acro{NB}{narrowband}
\acro{NBI}{narrowband interference}
\acro{NLA}{nonlinear sparse approximation}
\acro{NLOS}{Non-Line of Sight}
\acro{NTIA}{National Telecommunications and Information Administration}
\acro{NTP}{National Toxicology Program}
\acro{NHS}{National Health Service}

\acro{LOS}{Line of Sight}

\acro{OC}{optimum combining}                             
\acro{OC}{optimum combining}
\acro{ODE}{operational distortion-energy}
\acro{ODR}{operational distortion-rate}
\acro{OFDM}{Orthogonal Frequency-Division Multiplexing}
\acro{omp}[OMP]{orthogonal matching pursuit}
\acro{OSMP}[OSMP]{orthogonal subspace matching pursuit}
\acro{OQAM}{offset quadrature amplitude modulation}
\acro{OQPSK}{offset QPSK}
\acro{OFDMA}{Orthogonal Frequency-division Multiple Access}
\acro{OPEX}{Operating Expenditures}
\acro{OQPSK/PM}{OQPSK with phase modulation}

\acro{PAM}{pulse amplitude modulation}
\acro{PAR}{peak-to-average ratio}
\acrodef{pdf}[PDF]{probability density function}                      
\acro{PDF}{probability density function}
\acrodef{p.d.f.}[PDF]{probability distribution function}
\acro{PDP}{power dispersion profile}
\acro{PMF}{probability mass function}                             
\acrodef{p.m.f.}[PMF]{probability mass function}
\acro{PN}{pseudo-noise}
\acro{PPM}{pulse position modulation}
\acro{PRake}{Partial Rake}
\acro{PSD}{power spectral density}
\acro{PSEP}{pairwise synchronization error probability}
\acro{PSK}{phase shift keying}
\acro{PD}{power density}
\acro{8-PSK}[$8$-PSK]{$8$-phase shift keying}
\acro{PPP}{Poisson point process}
\acro{PCP}{Poisson cluster process}
 
\acro{FSK}{Frequency Shift Keying}

\acro{QAM}{Quadrature Amplitude Modulation}
\acro{QPSK}{Quadrature Phase Shift Keying}
\acro{OQPSK/PM}{OQPSK with phase modulator }

\acro{RD}[RD]{raw data}
\acro{RDL}{"random data limit"}
\acro{ric}[RIC]{restricted isometry constant}
\acro{rict}[RICt]{restricted isometry constant threshold}
\acro{rip}[RIP]{restricted isometry property}
\acro{ROC}{receiver operating characteristic}
\acro{rq}[RQ]{Raleigh quotient}
\acro{RS}[RS]{Reed-Solomon}
\acro{RSC}[RSSC]{RS based source coding}
\acro{r.v.}{random variable}                               
\acro{R.V.}{random vector}
\acro{RMS}{root mean square}
\acro{RFR}{radiofrequency radiation}
\acro{RIS}{Reconfigurable Intelligent Surface}
\acro{RNA}{RiboNucleic Acid}
\acro{RRM}{Radio Resource Management}
\acro{RUE}{reference user equipments}
\acro{RAT}{radio access technology}
\acro{RB}{resource block}

\acro{SA}[SA-Music]{subspace-augmented MUSIC with OSMP}
\acro{SC}{small cell}
\acro{SCBSES}[SCBSES]{Source Compression Based Syndrome Encoding Scheme}
\acro{SCM}{sample covariance matrix}
\acro{SEP}{symbol error probability}
\acro{SG}[SG]{sparse-land Gaussian model}
\acro{SIMO}{single-input multiple-output}
\acro{SINR}{signal-to-interference plus noise ratio}
\acro{SIR}{signal-to-interference ratio}
\acro{SISO}{Single-Input Single-Output}
\acro{SMV}{single measurement vector}
\acro{SNR}[\textrm{SNR}]{signal-to-noise ratio} 
\acro{sp}[SP]{subspace pursuit}
\acro{SS}{spread spectrum}
\acro{SW}{sync word}
\acro{SAR}{specific absorption rate}
\acro{SSB}{synchronization signal block}
\acro{SR}{shrink and realign}

\acro{tUAV}{tethered Unmanned Aerial Vehicle}
\acro{TBS}{terrestrial base station}

\acro{uUAV}{untethered Unmanned Aerial Vehicle}
\acro{PDF}{probability density functions}

\acro{PL}{path-loss}

\acro{TH}{time-hopping}
\acro{ToA}{time-of-arrival}
\acro{TR}{transmitted-reference}
\acro{TW}{Tracy-Widom}
\acro{TWDT}{TW Distribution Tail}
\acro{TCM}{trellis coded modulation}
\acro{TDD}{Time-Division Duplexing}
\acro{TDMA}{Time Division Multiple Access}
\acro{Tx}{average transmit}

\acro{UAV}{Unmanned Aerial Vehicle}
\acro{uric}[URIC]{upper restricted isometry constant}
\acro{urict}[URICt]{upper restricted isometry constant threshold}
\acro{UWB}{ultrawide band}
\acro{UWBcap}[UWB]{Ultrawide band}   
\acro{URLLC}{Ultra Reliable Low Latency Communications}
         
\acro{wuric}[UWRIC]{upper weak restricted isometry constant}
\acro{wurict}[UWRICt]{UWRIC threshold}                
\acro{UE}{User Equipment}
\acro{UL}{UpLink}

\acro{WiM}[WiM]{weigh-in-motion}
\acro{WLAN}{wireless local area network}
\acro{wm}[WM]{Wishart matrix}                               
\acroplural{wm}[WM]{Wishart matrices}
\acro{WMAN}{wireless metropolitan area network}
\acro{WPAN}{wireless personal area network}
\acro{wric}[WRIC]{weak restricted isometry constant}
\acro{wrict}[WRICt]{weak restricted isometry constant thresholds}
\acro{wrip}[WRIP]{weak restricted isometry property}
\acro{WSN}{wireless sensor network}                        
\acro{WSS}{Wide-Sense Stationary}
\acro{WHO}{World Health Organization}
\acro{Wi-Fi}{Wireless Fidelity}

\acro{sss}[SpaSoSEnc]{sparse source syndrome encoding}

\acro{VLC}{Visible Light Communication}
\acro{VPN}{Virtual Private Network} 
\acro{RF}{Radio Frequency}
\acro{FSO}{Free Space Optics}
\acro{IoST}{Internet of Space Things}

\acro{GSM}{Global System for Mobile Communications}
\acro{2G}{Second-generation cellular network}
\acro{3G}{Third-generation cellular network}
\acro{4G}{Fourth-generation cellular network}
\acro{5G}{Fifth-generation cellular network}	
\acro{gNB}{next-generation Node-B Base Station}
\acro{NR}{New Radio}
\acro{UMTS}{Universal Mobile Telecommunications Service}
\acro{LTE}{Long Term Evolution}

\acro{QoS}{Quality of Service}
\end{acronym}
	
	%% EMF definitions
\newcommand{\SAR} {\mathrm{SAR}}
\newcommand{\WBSAR} {\mathrm{SAR}_{\mathsf{WB}}}
\newcommand{\gSAR} {\mathrm{SAR}_{10\si{\gram}}}
\newcommand{\Sab} {S_{\mathsf{ab}}}
\newcommand{\Eavg} {E_{\mathsf{avg}}}
\newcommand{\ft}{f_{\textsf{th}}}
\newcommand{\alphatf}{\alpha_{24}}

\title{
Terrain-based Coverage Manifold Estimation: Machine Learning, Stochastic Geometry, or Simulation?
}
\author{
Ruibo Wang, Washim Uddin Mondal, Mustafa A. Kishk, {\em Member, IEEE}, Vaneet Aggarwal, {\em Senior Member, IEEE}, and Mohamed-Slim Alouini, {\em Fellow, IEEE}
\thanks{Ruibo Wang, Vaneet Aggarwal, and Mohamed-Slim Alouini are with King Abdullah University of Science and Technology (KAUST), CEMSE division, Thuwal 23955-6900, Saudi Arabia. Washim Uddin Mondal and Vaneet Aggarwal are with the Purdue University, West Lafayette, in 47907 USA. Mustafa A. Kishk is with the Department of Electronic Engineering, Maynooth University, Maynooth, W23 F2H6, Ireland. (e-mail: ruibo.wang@kaust.edu.sa; washim.uddinmondal@gmail.com; mustafa.kishk@mu.ie; vaneet@purdue.edu; slim.alouini@kaust.edu.sa). }

%\authorrefmark{1}
% \author{RUIBO WANG$^\mathrm{1}$, 
% WASHIM UDDIN MONDAL$^\mathrm{2}$, 
% MUSTAFA A. KISHK$^\mathrm{3}$ (Member IEEE), VANEET AGGARWAL$^\mathrm{2}$ (Senior Member IEEE), AND MOHAMED-SLIM ALOUINI$^\mathrm{1}$ \\ (Fellow IEEE)}
% \affil{King Abdullah University of Science and Technology (KAUST), CEMSE division, Thuwal 23955-6900, Saudi Arabia}
% \affil{Purdue University, West Lafayette, in 47907 USA}
% \affil{Department of Electronic Engineering, Maynooth University, Maynooth, W23 F2H6, Ireland}
% \corresp{CORRESPONDING AUTHOR: MUSTAFA A. KISHK (e-mail: mustafa.kishk@mu.ie).}
% %\authornote{This work was supported by the Natural Sciences and Engineering Research Council (NSERC) of Canada.}
% \markboth{Preparation of Papers for IEEE OPEN JOURNALS}{Author \textit{et al.}}
\vspace{-8mm}
}
\maketitle

\begin{abstract}
Given the necessity of connecting the unconnected, covering blind spots has emerged as a critical task in the next-generation wireless communication network. A direct solution involves obtaining a coverage manifold that visually showcases network coverage performance at each position. Our goal is to devise different methods that minimize the absolute error between the estimated coverage manifold and the actual coverage manifold (referred to as accuracy), while simultaneously maximizing the reduction in computational complexity (measured by computational latency). Simulation is a common method for acquiring coverage manifolds. Although accurate, it is computationally expensive, making it challenging to extend to large-scale networks. In this paper, we expedite traditional simulation methods by introducing a statistical model termed line-of-sight probability-based accelerated simulation. Stochastic geometry is suitable for evaluating the performance of large-scale networks, albeit in a coarse-grained manner. Therefore, we propose a second method wherein a model training approach is applied to the stochastic geometry framework to enhance accuracy and reduce complexity. Additionally, we propose a machine learning-based method that ensures both low complexity and high accuracy, albeit with a significant demand for the size and quality of the dataset. Furthermore, we describe the relationships between these three methods, compare their complexity and accuracy as performance verification, and discuss their application scenarios.
\end{abstract}

\begin{IEEEkeywords}
Coverage manifold, terrain, machine learning, stochastic geometry, simulation.
\end{IEEEkeywords}

\maketitle

\section{Introduction}
\IEEEPARstart{P}{roviding}  connectivity to the unconnected is a vital issue for next-generation networks \cite{yaacoub2020key}. To visually show the coverage-lacking locations, we can achieve a functional mapping that takes a network realization as input and produces its location-dependent coverage performance (referred to as the coverage manifold \cite{mondal2022deep}) as an output. In addition to showing coverage-lacking locations, the coverage manifold can provide references for coverage enhancement of future intelligent networks and temporary wireless network construction.

\par
The terrain is one of the most critical factors affecting the coverage manifold. When blocked by solid obstacles (i.e., buildings), the received signal propagates in a non-line-of-sight (NLoS) link and suffers severe attenuation. Compared to the unblocked line-of-sight (LoS) signal, the reception of NLoS signals is unstable, thus significantly influencing the coverage performance  \cite{thornburg2016performance}. The base station (BS)-user transmission in regions with high-rise buildings is even dominated by the NLoS propagation \cite{imran2019seamless}. Fortunately, obtaining the terrain topology in the real world is no longer difficult \cite{opencellid}. Therefore, considering the terrain when estimating the coverage manifold is feasible and necessary. As a result, this paper aims to find a method to accurately map the topology containing the location of BSs and buildings into a coverage manifold. In the following subsections, we present three potential mapping solutions and the challenges that need to be handled.

\subsection{Simulation}\label{introsimu}
Simulation is one of the widely used methods for coverage estimation. The simulation of coverage manifold under terrain has been carried out in the literature \cite{gesbert2022uav}. Simulation is intuitive and easy to interpret and does not need pre-training \cite{lou2021green}. However, estimating the coverage manifold based on terrain by simulation confronts challenges, especially in terms of computational complexity \cite{ozcan2020robust}. In the traditional simulation, the way to determine whether a BS-user link is blocked is to verify the blockage of all buildings, which is computationally expensive \cite{esrafilian20173d}. As a result, simulation has excellent advantages in coverage estimation in small areas because it is free of training and has high accuracy \cite{ozcan2020robust}. However, for the performance evaluation of large-scale wireless networks, the simulation method requires the support of strong computational power. In this paper, we provide a low-complexity simulation method that can be used for calculating coverage manifolds in large-scale networks.

\subsection{Stochastic Geometry (SG)}\label{introSG}
SG is a typical model-driven tool suitable for modeling and performance analysis of irregular network topology \cite{wang2022ultra}. Researchers in the field of SG have described the distributions of BSs' and users' locations by various models and accordingly have come up with analytical expressions for average coverage probability in a given area \cite{lou2023coverage}. Most of these analytical results are proved to be accurate in the symmetric scenarios \cite{wang2022evaluating}. A symmetric scenario is defined as a scenario where the point processes describing the locations of BSs and users, and the fading distribution of the channels are independent of each other \cite{alzenad2019coverage}. Under the assumption of the symmetry scenario, the authors in \cite{lou2023coverage} and \cite{wang2023resident} derive the analytical expression of coverage probability that considers the LoS and NLoS propagation. BSs and users follow homogeneous Poisson point processes (PPPs), while the channel experiences both shadowing and Nakagami-m small-scale fading \cite{lou2023haps}. 
\par

The process of SG-based coverage manifold generation can be broken down into two steps. Firstly, we extract some feature parameters from the network topology. Secondly, we substitute the parameters into the analytical expression of coverage probability and obtain the manifold. Typical input features include the density of BS, the height and the density of buildings, and so on \cite{al2014modeling}. When the BSs' and users' locations follow a homogeneous distribution, which is a common assumption under the SG framework, the distributions of BSs or users can be expressed by a single parameter called density \cite{andrews2016primer}. Because feature parameter extraction, BS locations, and terrain information are severely compressed, the coverage probability estimation is location-independent and the obtained coverage manifold is coarse-grained.

The complexity is not related to the number of buildings or BSs \cite{alzenad2019coverage} since these quantities are extracted as input features, such as the BS density. Therefore, the existing SG framework is only suitable for the average coverage performance evaluation of large-scale networks. Obtaining a location-dependent coverage manifold is beyond the reach of existing SG-based methods.

\subsection{Machine Learning (ML)}\label{introML}
Applying machine learning in wireless communication networks provides adaptability to dynamic environments \cite{sun2019application,wu2023efficient,tang2023active,wu2020efficient}, which in turn enables intelligent resource optimization \cite{chang2018machine,sun2018deep}. This intelligent resource allocation leads to efficient resource utilization and enables the provision of personalized services \cite{chang2018joint}. These combined advantages ultimately enhance both network performance and the overall user experience \cite{fei2016survey}.

\par
ML and SG are compared in several studies in the field of wireless communication. We conclude that there are three main differences between them. Firstly, SG and ML are typical representatives of model-driven and data-driven methods, respectively \cite{hmamouche2021new}. Specifically, SG learns from previous scholars' models and studies, while ML learns from data sets. Secondly, the SG analytical framework is interpretable because the final results, in most cases, are analytical functions of the input features. ML-based methods, on the other hand, provide more accurate results at the cost of reduced interpretability. Finally, another feature of ML-based methods is that unlike SG-based methods \cite{el2018machine}, they obliterate the need for feature extraction \cite{mondal2022deep}.

\par
In recent years, there have been several studies that applied ML-based methods to the evaluation of various system-level parameters (such as data rate \cite{mondal2022deep}, coverage probability \cite{mondal2022deep,el2018machine} and energy efficiency \cite{zappone2019model}). 
%Especially, our previous work \cite{mondal2022deep} is the first work in ML-based coverage manifold estimation. 
These studies all prove that the ML-based method has low complexity and high evaluation accuracy. In this paper, we model the coverage probability of a reference point as a function of the height distribution of nearby buildings and realize low-complexity coverage manifolds estimation. Compared with the above work, our proposed ML approach is significantly different. Firstly, the proposed ML approach takes the terrain as one of the inputs for the first time, which makes the coverage manifold estimation much more challenging. Secondly, in the above studies, the ML-based method is the main body, and SG/simulation serves as a baseline or assistance. In our research, ML, SG, and simulation-based methods are on equal status and get a relatively fair comparison. Thirdly, our proposed method is interpretable and analyzable in complexity.

\subsection{Contribution}
To our best knowledge, this is the first work on terrain-based coverage manifold estimation. Our main contributions are given as follows. Firstly, we propose three methods for estimating coverage manifold. 
\begin{itemize}
    \item We accelerated the traditional simulation by reducing the number of rounds for blockage verification and the complexity of one round blockage verification. The accelerated simulation, with minimal loss in accuracy, exhibits a significant reduction in computational complexity.
    \item We simplify and adapt the existing SG analytical results for coverage manifolds estimation. We propose a model training method to reduce the computational complexity. As a result, the SG-based method is fast but coarse-grained. 
    \item We propose a data-driven method based on ML. By inputting the gridded positions and heights of BSs and buildings, the trained model can output high-accuracy coverage manifold with relatively low complexity.
\end{itemize}
Secondly, with regard to coverage manifold estimation, we establish the connections and present the differences among these three methods. 
\begin{itemize}
    \item \textbf{Differences between three methods:} Through the analysis of their computational complexity and comparison of accuracy and time delay, these three methods sacrifice different degrees of accuracy in exchange for different computational gains. Furthermore, the different application scenarios suitable for each method are provided in the conclusion.
    \item \textbf{The connection between accelerated simulation and SG:} With the aid of a statistical blockage model called the air-to-ground LoS probability model (A2GLPM), researchers have derived analytical expressions for terrain-based coverage probability under the SG analysis framework. We use A2GLPM to accelerate traditional simulation.
    \item \textbf{The connection and  difference between ML and SG:} Referring to the concept of linear regression, we design both the ML-based method and the SG-based method. SG and ML are used for model-driven and data-driven tasks, respectively. Therefore, we employ different training approaches, based on models and datasets, to train them.
\end{itemize}
Finally, Table~\ref{acronyms} and Table~\ref{variables} summarize the main acronyms and main variables in this article with their meaning, respectively.

\begin{table}[ht]
\centering
\caption{Summary of the main acronyms and their meaning.}
\label{acronyms}
\resizebox{0.45\linewidth}{!}{
\renewcommand{\arraystretch}{1.1}
\setlength{\tabcolsep}{0.3mm}{
\begin{tabular}{|c|c|}
\hline
Acronym    & Meaning   \\ \hline
LoS   & Line-of-sight   \\ \hline
NLoS     & Non-line-of-sight     \\ \hline
BS & Base station   \\ \hline
SG   & Stochastic geometry \\ \hline
PPP & Poisson point process  \\ \hline
ML & Machine learning  \\ \hline
A2GLPM &  Air-to-ground LoS probability model \\ \hline
PDF & Probability density function  \\ \hline
MRP &  Manifold receive position  \\ \hline
\end{tabular}}
}
\end{table}

\begin{table}[ht]
\centering
\caption{Summary of the main variables and their meaning.}
\label{variables}
\resizebox{0.7\linewidth}{!}{
\renewcommand{\arraystretch}{1.1}
\setlength{\tabcolsep}{0.3mm}{
\begin{tabular}{|c|c|}
\hline
Variable   & Meaning   \\ \hline
$N$   & Number of BSs   \\ \hline
$M$  & Number of buildings    \\ \hline
$\Phi_i$   & Latitude and longitude of the $i^{th}$ building's endpoints \\ \hline
$n_i$   & Number of the $i^{th}$ building's endpoints \\ \hline
$\left(\overline{x}_i, \overline{y}_i \right)$ & Center of the $i^{th}$ building  \\ \hline
$r_i$ & Radius of the smallest enclosing circle of the $i^{th}$ building \\ \hline
$h_i$ &  Height distribution of the $i^{th}$ building  \\ \hline
$\left( x_{\rm{MRP}}, y_{\rm{MRP}} \right)$ & Coordinates of the typical MRP \\ \hline
$R_{\odot}$ & Horizontal radius of the terrain information neighborhood   \\ \hline
$P_{\mathrm{LoS}}$ &  Probability of the $i^{th}$ BS-MRP link is LoS \\ \hline
$h_{\mathrm{BS}}$ &  Height of BSs  \\ \hline
$d_i$ &  Distance between the $i^{th}$ BS and MRP \\ \hline
$P^C$ &  Coverage probability \\ \hline
$B$ &  Matrix recording the positions of BSs  \\ \hline
$H$ &  Matrix recording the positions and heights of buildings  \\ \hline
$d_{\mathrm{TH}}$ &  Threshold that determines whether the A2GLPM is applied  \\ \hline
\end{tabular}}
}
\end{table}

\section{System Model}
In this section, we establish the terrain model, BS distribution model, and channel model. These models are inputs of simulation, SG, or ML methods. Next, an essential statistical model called A2GLPM is introduced. The A2GLPM is not part of the inputs but is involved in generating the coverage manifold. Finally, we introduce the data sets for training and testing.

\subsection{Terrain and BS Models}\label{sec2-1}
\begin{figure}[t]
\centering
\includegraphics[width=0.7\linewidth]{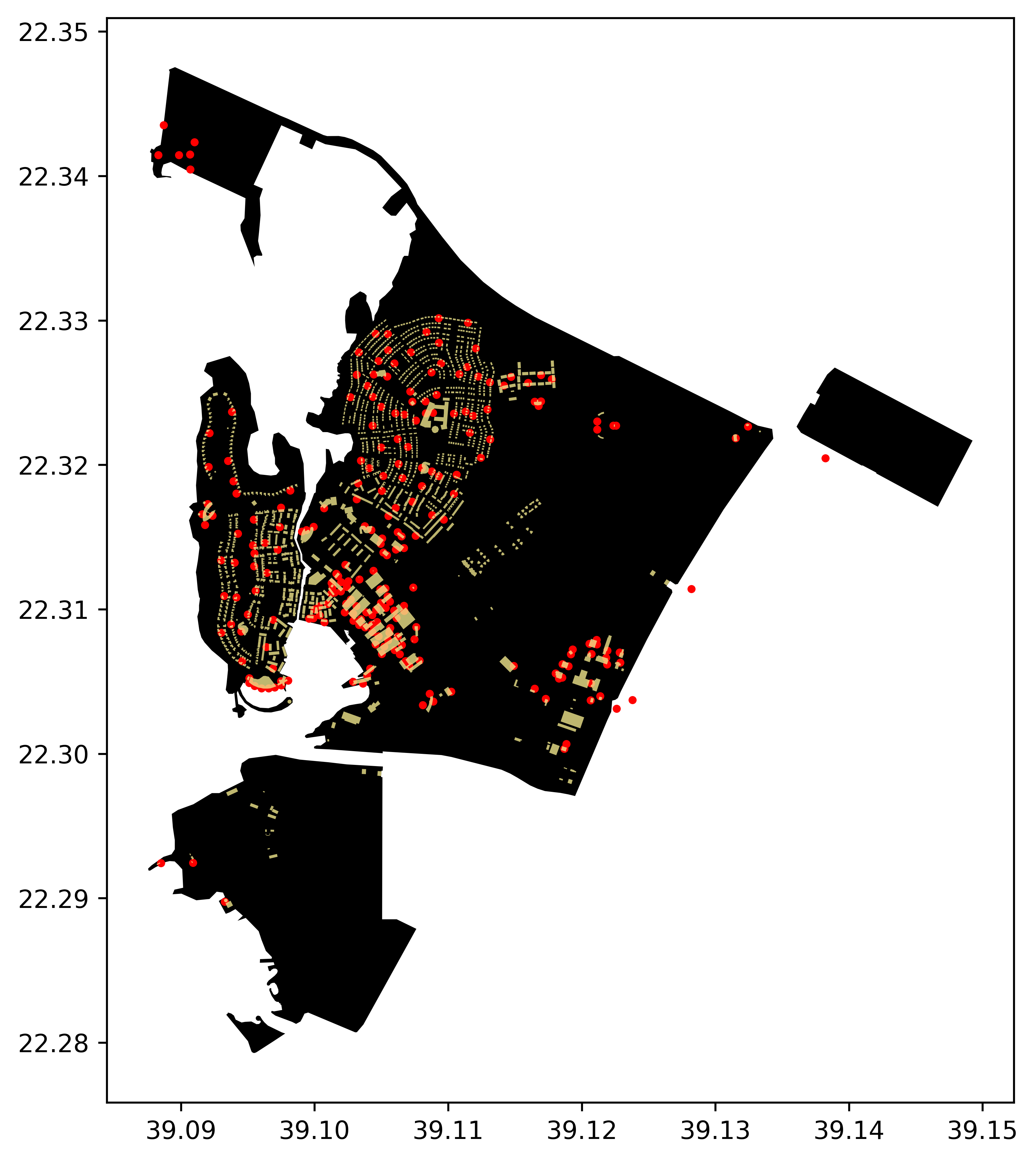}
\caption{The terrain of KAUST ($M=2229$, $N=243$ and $\omega=9$).}
\label{figure1}
\end{figure}

Denote the number of BSs as $N$ and the longitude and latitude of the $i^{th}$ BS as $\left( x_{{\rm{BS}},i} , y_{{\rm{BS}},i} \right)$, $i \leq N$. \footnote{The latitude and longitude of the BS location can be downloaded from the website www.opencellid.org. This paper also converts them into coordinates in the Cartesian coordinate system.} The heights of all BS are assumed to be the same for the convenience of generating analytical results. In particular, the height of a BS $h_{\rm{BS}}$ is taken to be 20m. In Fig.~\ref{figure1}, the red dots are the locations of the BSs while the yellow polygons denote the floor areas of the buildings.

\par
Given that there are $M$ buildings in a particular area, and the terrain information of the $i^{th}$ building can be represented by the set $\mathcal{B}_i = \left\{ \Phi_i,  \left(\overline{x}_i, \overline{y}_i \right), r_i, h_i, \mathcal{A}_i \right\}$  for the which we provide detailed explanations next. We assume that the $i^{th}$ building has $n_i-1$ endpoints whose cartesian coordinates are given as $\Phi_i = \left\{ \left(x_i^1, y_i^1 \right), \left(x_i^2, y_i^2 \right),  \dots \left(x_i^{n_i}, y_i^{n_i} \right) \right\}$ in clockwise order\footnote{The dataset recording the latitude and longitude of the buildings' endpoints can be downloaded from the website www.openstreetmap.org. Since the given area is small, we ignore the spherical effect of the Earth. Remote areas with sparse architectural distribution, such as mountains and lakes, are not considered typical samples.} Latitude and longitude are converted to coordinates in the Cartesian coordinate system. Every two adjacent endpoints of the set $\Phi_i$, such as $\left(x_i^1, y_i^1 \right)$ and $\left(x_i^2, y_i^2 \right)$, form an edge of the building. To ensure that the geometry formed by the endpoints (shown as the yellow contour in Fig.~\ref{figure1}) is closed, $x_i^1=x_i^{n_i}$ and $y_i^1=y_i^{n_i}$ need to be satisfied for any $i \leq M$. $\left(\overline{x}_i, \overline{y}_i \right)$ and $r_i$ are the center and radius of the smallest enclosing circle of the $i^{th}$ building. The solution for the minimum bounding circle can be found in \cite{xu2003solution}. $\mathcal{A}_i$ records the floor area of the building, which can be calculated by the Shoelace Theorem \cite{tables1987boca}. Furthermore, we assume the building's height distribution $h_i$ follows the Rayleigh probability density function (PDF): $f_H(h_i) = \left(h_i / \omega^2 \right) \exp \left( -h_i^2 / 2\omega^2 \right)$, which is a common assumption about the building height distribution \cite{al2014optimal}, and $\omega$ is called the scale parameter of the Rayleigh distribution.

\par
It is clear from the above discussion that the goal of data processing is primarily to obtain the center $\left(\overline{x}_i, \overline{y}_i \right)$ and radius $r_i$ of the minimum bounding circle of each building. Considering that the computational complexity of calculating $\left(\overline{x}_i, \overline{y}_i \right)$ and $r_i$ for a single building is only $\mathcal{O} \left( n_i \right)$ \cite{xu2003solution}, the computational complexity of data pre-processing is ignored in the following sections.

\subsection{Channel Model}
Our main goal is to obtain downlink coverage probability at all those points where a mobile user could be potentially located. In this paper, such potential locations are referred to as the manifold receive positions (MRPs). Although, in general, the set of MRPs could be continuous, for computational tractability, we assume it to be a collection of discrete points. In the outdoor public wireless network coverage scenario, the MRPs are assumed to be on the ground with a height of 0m. The channel between a BS and an MRP experiences shadowing $\eta_Q$ and small-scale fading $G_{Q,i}$, where $Q$ is replaced by $\rm{LoS}$ when there are no buildings blocking the link between the MRP and the BS; otherwise, $Q$ is replaced by $\rm{NLoS}$. Since the process of obtaining coverage performance is the same at all MRPs, below we explain the process only for a typical MRP located at coordinates $\left( x_{\rm{MRP}}, y_{\rm{MRP}} \right)$. Note that the power received at this point from the $i^{th}$ BS can be expressed as,
\begin{equation}
    S_{Q,i} = \rho \, \eta_Q \, G_{Q,i} \, d_{i}^{\alpha_Q},
\end{equation}
where $G_{Q,i}$ follows the Nakagami-m fading experienced with shape parameters and scale parameters $\left( m_Q, \frac{1}{m_Q}\right)$, and the PDF of $G_{Q,i}$ is given by \cite{nakagami}
\begin{equation}
    {f_{G_{Q,i}}}\left(g\right) = \frac{{{m_Q}^{{m_Q}}{g^{{m_Q} - 1}}}}{{\Gamma \left( {{m_Q}} \right)}}{e^{ - {m_Q} \, g}},
\end{equation}
where $\Gamma \left( {{m_Q}} \right) = \int_0^\infty  {{z^{{m_Q} - 1}}{e^{ - z}}dz}$ is the Gamma function, $d_i$ is the distance between the $i^{th}$ BS and the MRP,
\begin{equation}\label{dij}
    d_i = \sqrt{ \left(x_{{\rm{BS}},i} - x_{\rm{MRP}}\right)^2 + \left(y_{\rm{BS}} - y_{\rm{MRP}}\right)^2 + h_{\rm{BS}}^2 }.
\end{equation}

Table~\ref{table1} describes the definitions and values of other parameters \cite{alzenad2019coverage}.

\begin{table}[H]
\centering
\caption{Summary of channel parameters.}
\label{table1}
\renewcommand{\arraystretch}{1.1}
\setlength{\tabcolsep}{0.25mm}{
\begin{tabular}{|c|c|c|}
\hline
Parameter                                  & Symbol                & Value        \\ \hline
Transmission power                     & $\rho$                   & 30~dBm       \\ \hline
Bandwidth                                  & $B$                   & 10~MHz       \\ \hline
Environmental noise power spectral density & $\sigma$            & -174~dBm/Hz  \\ \hline
SINR threshold                             & $\gamma$              & 0~dB         \\ \hline
LoS/NLoS additional loss                   & $\eta_{\rm{LoS}}, \eta_{\rm{NLoS}}$      & -38.6, -59.5~dB \\ \hline
LoS/NLoS path loss exponent                & $\alpha_{\rm{LoS}}, \alpha_{\rm{NLoS}}$ & 2, 3         \\ \hline
LoS/NLoS Nakagami-m fading parameter & $m_{\rm{LoS}}, m_{\rm{NLoS}}$            & 2, 1         \\ \hline
\end{tabular}
}
\end{table}

\subsection{A2GLPM}
The A2GLPM is a statistical model capturing the relative position between BS and MRP and describing the probability of being blocked by buildings \cite{al2014optimal}. This model plays a key role in the design of the three mapping methods. To a certain extent, this model can replace the frequent verification of building blockage in simulation and significantly reduce the computational complexity of the simulation. Furthermore, the A2GLPM has been proven to fit well into the SG framework \cite{wang2018modeling} and is the core concept for the design of ML methods.

\par
The A2GLPM establishes the relation between the BS-MRP elevation angle and the probability of not being blocked. The tangent of the elevation angle is given as the ratio of the height of the BS to the horizontal distance between the BS and the MRP. The probability of the $i^{th}$ BS and the typical MRP establishing an LoS link is \cite{al2014optimal},
\begin{equation}\label{PLoSk}
\begin{split}
    & P_{\rm{LoS}} \left( a,b,d_i \right) \\
    & = \frac{1}{1 + a\exp \left( { - b\left( {\frac{{180}}{\pi }{{\tan }^{ - 1}}\left( {\frac{h_{\rm{BS}}}{\sqrt{d_i^2-h_{\rm{BS}}^2}}} \right){-a}} \right)} \right)},
%\vspace{-0.15cm}
\end{split}
\end{equation}
where $d_i$ is the distance between the BS and MRP defined in (\ref{dij}), $\frac{h_{\rm{BS}}}{\sqrt{d_i^2-h_{\rm{BS}}^2}}$ is the tangent of the elevation angle. $a$ and $b$ are terrain-related parameters, which can be calculated by three terrain parameters:
\begin{equation}\label{a and b}
    Q = \sum_{j=0}^3 \sum_{i=0}^{3-j} C_{i,j} \left(\kappa \iota \right)^i \omega^j,
\end{equation}
where $Q \in \{a,b\}$, $C_{i,j}$ are polynomial coefficients. The values of $C_{i,j}$ can refer to Table \uppercase\expandafter{\romannumeral1} and Table \uppercase\expandafter{\romannumeral2} in \cite{al2014optimal}. The physical meanings of $\kappa$, $\iota$ and $\omega$ are as follows:
\begin{itemize}
    \item $\kappa$ represents the ratio of the built area to the total area (dimensionless).
    \item $\iota$ represents the mean number of buildings per unit area (buildings/km$^2$).
    \item $\omega$ is the scale parameter of Rayleigh distribution that describes the buildings' heights (dimensionless and introduced in Sec.~\ref{sec2-1}).
\end{itemize}

\par
Finally, based on basic geometric relationships, it is easy to know that the closer a building is to the MRP, the more likely it is to block the MRP-BS link. Therefore, we assume that parameters $a$ and $b$ only focus on the terrain information in the circular neighborhood centered at the MRP with horizontal radius $R_{\odot}$. Therefore, $\kappa$ and $\iota$ of the MRP consider the terrain information in its circular neighborhood, and the edge effect is ignored. As long as the center of the $i^{th}$ building $\left(\overline{x}_i, \overline{y}_i \right)$ is located in the circular neighborhood, its whole floor area $\mathcal{A}_i$ is considered inside the neighborhood.

\section{Accelerated Simulation}\label{section3}
Traditional simulation is computationally expensive. It needs to verify whether each BS-MRP link is blocked, and each verification goes through the buildings' top edges. For a simple scenario with $N$ BSs and $M$ $n$-prism buildings, traditional simulation requires $NMn$ rounds of blockage verification. To accelerate the simulation, we reduce the computational load of blockage verification by leveraging geometric relationships for BSs closer to the user and we employ A2GLPM instead of blockage verification for BSs farther away from the user.

\subsection{Blockage Verification}
This subsection proposes algorithms that determine the building blockage of the BS-MRP link. Blockage can be divided into two types, which are defined as follows.

\begin{definition}[Type-I blockage]
    When MRP is located inside one of the buildings, the resulting building blockage is called type-I blockage.
\end{definition}
\begin{definition}[Type-II blockage]
    When the MRP is located outside all buildings and the MRP-BS link is blocked by at least one building, the resulting building blockage is called type-II blockage.
\end{definition}

\begin{algorithm}[!ht] 
	\caption{Type-I Blockage Verification Algorithm}
	\label{Alg1}
	\begin{algorithmic} [1]
		\FOR{$j = 1 : M$}

        \IF {$\left(\overline{x}_j - x_{\rm{MRP}}\right)^2 + \left(\overline{y}_j - y_{\rm{MRP}}\right)^2 < r_j^2$} \label{step1-3}    
        \STATE $n_{{\rm{PoI}},j}\leftarrow 0$, $A_1 \leftarrow \frac{y_j^k-y_j^{k+1}}{x_j^{k+1}-x_j^k}$, $B_1 \leftarrow \frac{y_j^{k+1}-y_j^k}{x_j^{k+1}-x_j^k} x_j^k - y_j^k$.
        
        \FOR{$k = 1 : n_j - 1$}
        \IF {$\frac{y_{\rm{MRP}}-B_1}{A_1}>x_{\rm{MRP}}$ \textbf{and} $\left( \frac{y_{\rm{MRP}}-B_1}{A_1} -x_j^k \right) \left( \frac{y_{\rm{MRP}}-B_1}{A_1} -x_j^{k+1} \right) <0$}
        \STATE $n_{{\rm{PoI}},j} \leftarrow n_{{\rm{PoI}},j}+1$. 
        \ENDIF
        \ENDFOR
        
        \IF {$n_{{\rm{PoI}},j}$ is odd}
        \STATE \textbf{Output} $V^{(1)} \leftarrow 1$ and \textbf{exit} the type-I blockage verification algorithm.
        \ENDIF 
        
        \ENDIF
		\ENDFOR
		\STATE \textbf{Output} $V^{(1)} \leftarrow 0$.
	\end{algorithmic}
\end{algorithm}

\begin{figure}[t]
\centering
\includegraphics[width=0.5\linewidth]{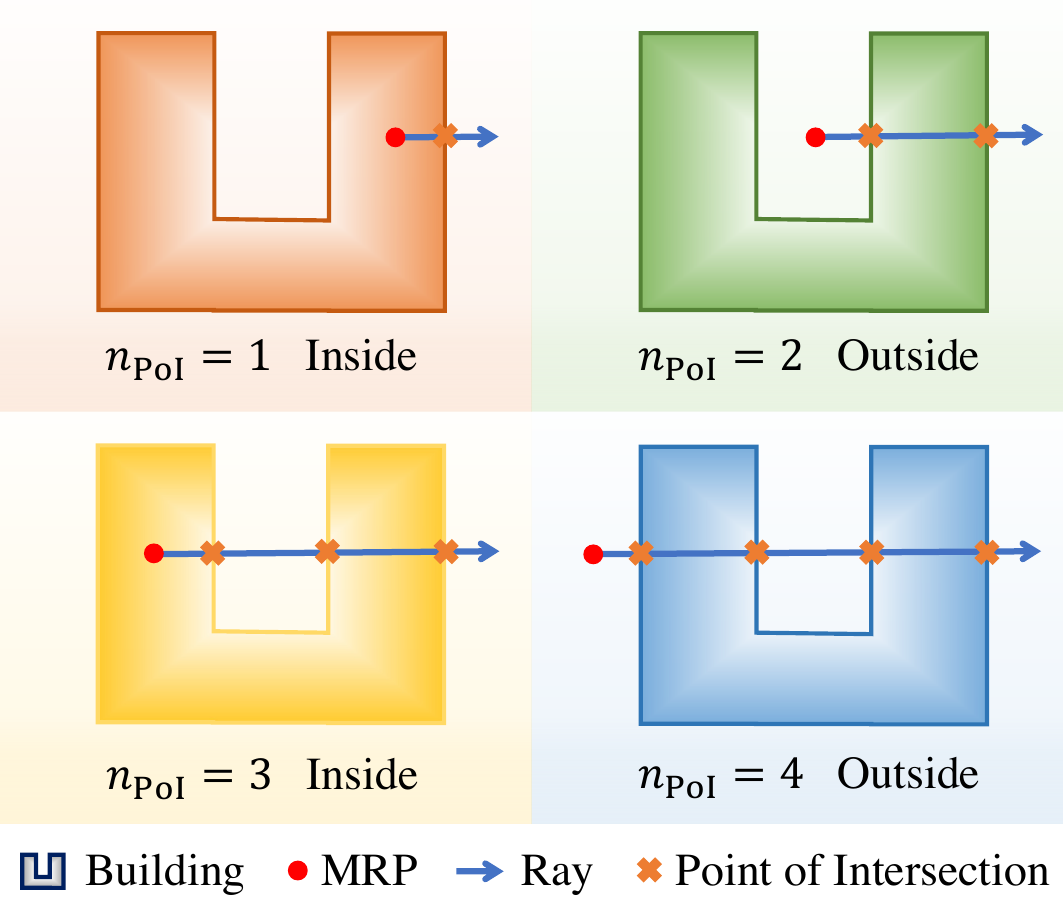}
\caption{Diagram of the right ray validation.}
\label{figure2}
\end{figure}

In the type-I blockage verification algorithm, if MRP locates inside any building, $V^{(1)}=1$, otherwise $V^{(1)}=0$. Step (3) in algorithm~\ref{Alg1} verifies whether the MRP locates in the smallest enclosing circle of the $i^{th}$ building. This verification dramatically reduces the complexity of algorithm~\ref{Alg1} because steps (4)-(13) only require one round of execution in most cases. Step (6) determines whether there is an intersection point between the rays extending to the right starting from MRP and a side of the building, and $n_{{\rm{PoI}},j}$ records the number of points of intersection (PoI) between the right ray of MRP and the $j^{th}$ building. The reason for verifying the type-I blockage through the parity of $n_{{\rm{PoI}},j}$ can be seen in Fig.~\ref{figure2}. The building shape in Fig.~\ref{figure2} refers to a building whose coordinate is around (39.11, 22.31) in Fig.~\ref{Alg1}.

\begin{algorithm}[!ht] 
	\caption{Type-II Blockage Verification Algorithm for the $i^{th}$ BS}
	\label{Alg2}
	\begin{algorithmic} [1]
        \STATE $A_2 \leftarrow \frac{y_{{\rm{BS}},i}-y_{\rm{MRP}}}{x_{\rm{MRP}}-x_{{\rm{BS}},i}}$, $B_2 \leftarrow \frac{y_{{\rm{BS}},i}-y_{\rm{MRP}}}{x_{{\rm{BS}},i} - x_{\rm{MRP}}} x_{{\rm{BS}},i} - y_{{\rm{BS}},i}$.
        
        \FOR{$j = 1 : M$}
        
        \IF{$ \frac{\left| A_2 \overline{x}_j - \overline{y}_j - B_2 \right|}{\sqrt{1+A_2^2}} < r_i$}
        
        \FOR{$k = 1 : n_j - 1$}

        \STATE $A_1 \leftarrow \frac{y_j^k-y_j^{k+1}}{x_j^{k+1}-x_j^k}$, $B_1 \leftarrow \frac{y_j^{k+1}-y_j^k}{x_j^{k+1}-x_j^k} x_j^k - y_j^k$.
        
        \IF {$\frac{h_{{\rm{BS}},i}}{h_{{\rm{BS}},i}-x_{\rm{MRP}}} \left( \frac{B_2-B_1}{A_1-A_2} - x_{\rm{MRP}} \right)<h_j$} 
        \STATE $V_i^{(2)} \leftarrow 1$ and \textbf{exit} the type-II blockage verification algorithm.
        \ENDIF
		\ENDFOR
        \ENDIF
        \ENDFOR
		\STATE \textbf{Output $V_i^{(2)} \leftarrow 0$}.
  
	\end{algorithmic}
\end{algorithm}	

\begin{figure}[t]
\centering
\includegraphics[width=0.7\linewidth]{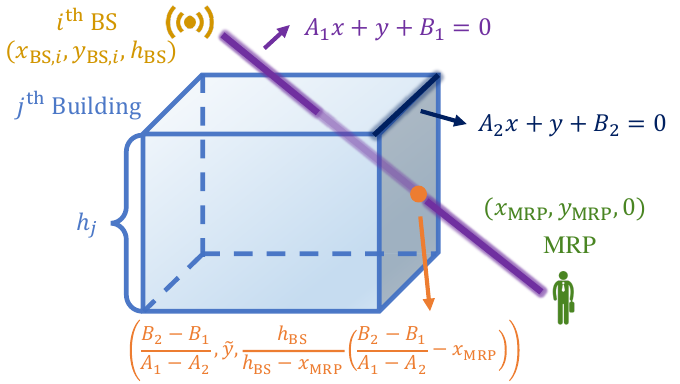}
\caption{Diagram of blockage verification.}
\label{figure3}
\end{figure}

In the type-II blockage verification algorithm, if the $i^{th}$ BS-MRP link is blocked by any building, $V_i^{(2)}=1$, otherwise $V_i^{(1)}=0$. Step (3) in algorithm~\ref{Alg2} verifies whether the line corresponding to the BS-MRP passes through the smallest enclosing circle of the $i^{th}$ building. Step (6) determines whether the building blocks the MRP-BS link by calculating the height of the intersection point between the building and the link. Fig.~\ref{figure3} is a schematic illustration explaining step (6).

\subsection{Accelerated Simulation Algorithm}
The accelerated simulation algorithm can be roughly divided into the following parts: The first part is blockage verification, corresponding to steps (1)-(14) in algorithm~\ref{Alg3}; The second part is applying the Monte Carlo method in solving coverage probability, corresponding to steps (15)-(26) in algorithm~\ref{Alg3}.

\begin{algorithm}[!ht] 
	\caption{Accelerated Simulation Algorithm}
	\label{Alg3}
	\begin{algorithmic} [1]
        \STATE $P^C_{\rm{AcSimu}} \leftarrow 0$. Run the type-I blockage verification algorithm. \textbf{If} $V^{(1)}=1$, \textbf{output} $P^C_{\rm{AcSimu}}=0$ and \textbf{exit} the accelerated simulation algorithm.
        \STATE $\kappa \leftarrow 0$, $\iota \leftarrow 0$.
        \FOR{$j = 1 : M$}
        \IF{$\left(\overline{x}_j - x_{\rm{MRP}} \right)^2 + \left(\overline{y}_j - y_{\rm{MRP}} \right)^2 < R_\odot^2$}
        \STATE $\kappa \leftarrow \kappa + \frac{\mathcal{A}_j}{\pi R_\odot^2}$ , $\iota \leftarrow \iota + \frac{1}{\pi R_\odot^2}$.
        \ENDIF
        \ENDFOR
        \STATE Obtain $a$ and $b$ through $\kappa$, $\iota$ and $\omega$ according to equation (\ref{a and b}).
        
        \FOR{$i = 1 : N$}
        \IF{$d_i>d_{\rm{TH}}$}
        \STATE $V_i^{(2)} \leftarrow \mathbbm{I} \left\{ {\mathbf{\rm{rand}}}(0,1) > P_{\rm{LoS}}\left(a,b,d_i\right) \right\}$.
        \ELSE
        \STATE Run the type-II blockage verification algorithm for the $i^{th}$ BS and obtain $V_i^{(2)}$.
        \ENDIF
        \STATE $\overline{S}_i \leftarrow  \left( 1-V_i^{(2)} \right) T \eta_{\rm{LoS}} d_i^{\alpha_{\rm{LoS}}} + V_i^{(2)} T \eta_{\rm{NLoS}} d_i^{\alpha_{\rm{NLoS}}}$.
        \ENDFOR
        %\left( 1-V_i^{(2)} \right) \mathbbm{E}_{G_{\rm{LoS}}} \left[ S_{{\rm{LoS}},i} \right] + V_i^{(2)} \mathbbm{E}_{G_{\rm{NLoS}}} =

        \STATE $i^* \leftarrow \underset{i}{\argmax} \ S_{{\rm{LoS}},i} $.

        \FOR{$k = 1 : n_{\rm{iter}}$}
        \FOR{$i = 1 : N$}
        \STATE Update $G_{Q,i}$ and $S_i \leftarrow \overline{S}_i G_{Q,i}$, where $Q \leftarrow {\rm{LoS}}$ for $V_i^{(2)}=0$, otherwise $Q \leftarrow {\rm{NLoS}}$.
        \ENDFOR
        \IF{${\rm{SINR}} = \frac{S_{i^*}}{\sum_{i=1}^N S_i - S_{i^*} + \sigma B}>\gamma$}
        \STATE $P^C_{\rm{AcSimu}} \leftarrow P^C_{\rm{AcSimu}} + \frac{1}{n_{\rm{iter}}}$.
        \ENDIF
        \ENDFOR
        \STATE \textbf{Output $P^C_{\rm{AcSimu}}$}.
	\end{algorithmic}
\end{algorithm}	

For the blockage verification part, when the MRP is located inside the building, the links to all the BSs are NLoS.  Steps (2)-(8) in algorithm~\ref{Alg3} are aiming at obtaining terrain-related parameters $a$ and $b$ in the A2GLPM. We perform the accurate but highly complex algorithm~\ref{Alg2} for the BS with a relatively close distance, that is, less than threshold $d_{\rm{TH}}$. Note that $d_{\rm{TH}}$ is a horizontal distance threshold. For distant BSs, the A2GLPM is applied to verify blockage, thus greatly reducing complexity. In step (11), $\mathbbm{I}\{ \cdot \}$ is an indicator function. When the condition $\{ \cdot \}$ is satisfied, $\mathbbm{I}\{ \cdot \}=1$, otherwise $\mathbbm{I}\{ \cdot \}=0$. ${\rm{rand}}(0,1)$ is a random number uniformly distributed between 0 and 1. $n_{\rm{iter}}$ in step (18) is the number of iterations for Monte Carlo simulation.
\par

For the coverage probability solving part, step (17) follows the strongest average received power association strategy \cite{mondal2020exact}. The sum of the power received at the MRP from other BSs is interference. ${\rm{SINR}}$ is a random variable due to the influence of small-scale fading. Therefore, we repeatedly update the small-scale fading and iterative ${\rm{SINR}}$ in steps (18)-(25) in algorithm~\ref{Alg3}. Then, the proportion of rounds in which ${\rm{SINR}}$ is greater than the threshold $\gamma$ is recorded as the coverage probability of simulation.

\subsection{Complexity Analysis}\label{section3-3}
In this subsection, we delve into the accelerated simulation's benefits in computational complexity. Examining the average complexity is more pertinent than focusing on the upper limit, as both the accelerated and traditional simulations involve $NMn$ rounds of blockage verification in the worst-case scenario. Hence, this paper refers to average complexity, unless specified otherwise.

\par
As mentioned, the probability of the typical MRP locating into the smallest enclosing circles of multiple buildings is low. On average, the complexity of the type-I blockage verification is only $\mathcal{O}(M)$. The complexity of verifying by the A2GLPM is much lower than that of type-II blockage in algorithm~\ref{Alg2}. In step (13) algorithm~\ref{Alg3}, there are about $\lambda \pi d_{\rm{TH}}^2$ BSs need to perform algorithm~\ref{Alg2}, and no more than $2 \iota \, d_{\rm{TH}} \, \overline{r}$ buildings that need to perform blockage verification on each side, where $\overline{r}=\frac{1}{M} \sum_i^M r_i$. Therefore, the complexity of the type-II blockage verification is reduced from $\mathcal{O}(NM\, \overline{n})$ of the traditional simulation to the current $\mathcal{O}(\pi \lambda \, \iota \, d_{\rm{TH}}^3 \overline{r} \,\overline{n})$, where $\overline{n}=\frac{1}{M} \sum_i^M n_i$. The complexity of coverage probability solving is mainly determined by updating the small-scale fading in step (20). In general, thousands of iterations are enough to meet the accuracy requirements, close to $M$ on the order of magnitude. Although the total round of channel realizations in step (20) is $N\,n_{\rm{iter}}$, which is much larger than the rounds of type-II blockage verification in accelerated simulation, the unit complexity in steps (3)-(10) of algorithm~\ref{Alg2} is much higher than that of step (20) in algorithm~\ref{Alg3}. Therefore, through numerical testing, the complexity comparison results are as follows:
\begin{itemize}
    \item In typical scenarios, the simulation's complexity is primarily determined by the rounds of type-II blockage verification. Using the verification of one side of the building as the unit complexity, the accelerated simulation can reduce the complexity from $\mathcal{O}(NM\overline{n})$ to $\mathcal{O}( \pi \lambda \, \iota \, d_{\rm{TH}}^3 \overline{r} \,\overline{n})$. 
    \item 
    In special cases where there are few buildings ($M$ is small) and a high demand for coverage accuracy ($n_{\rm{iter}}$ is large), we consider the small-scale fading updating as the unit complexity. In this scenario, the complexity of both the accelerated simulation and the traditional simulation is $\mathcal{O}(N\,n_{\rm{iter}})$.
    \item To get the coverage probability of the typical MRP in KAUST (normal case), the traditional simulation requires $2.67\times 10^6$ rounds of type-II blockage verification, while the accelerated simulation only needs $\{ 11.5, 40.6, 276.6, 3568.7\}$ rounds on average for $d_{\rm{TH}}=\{200, 300, 500, 1000 \}$ meters.
\end{itemize}

\section{SG-Based Method}
In this section, we design a method to apply the SG framework to obtain coverage manifold and analyze this method's complexity and accuracy. Before this, a model training procedure is proposed to reduce the complexity of the SG-based method.

\subsection{Coverage Probability Expression}
Several SG-based studies have derived terrain-related analytical expressions of coverage probability for different BS distributions. After weighing complexity and accuracy, we assume that the MRP only selects an LoS BS to provide service to simplify the expression. The approximate analytical expression of coverage probability is given as follows \cite{alzenad2019coverage},
\begin{equation}
\begin{split}\label{SGcoverage}
    & P^C \left(a,b,\lambda \right) \\
    & = \sum_{k=1}^{m_{\rm{LoS}}} \binom{m_{\rm{LoS}}}{k} \left(-1\right)^{k+1} \int_{h_{\rm{BS}}}^{R_{\max}} f_{\rm{LoS}} \left( a,b,\lambda,r \right) \\
    & \times \mathcal{L} \left( a,b,\lambda, k \left(m_{\rm{LoS}} !\right)^{\frac{-1}{m_{\rm{LoS}}}} m_{\rm{LoS}} \gamma \, \rho^{-1} \, \eta_{\rm{LoS}}^{-1} r^{\alpha_{\rm{LoS}}} \right) \mathsf{d} r,
\end{split}
\end{equation}
where $R_{\max}$ is the maximum distance that the MRP can receive signals from the BS, and $\lambda$ is the density of BS, which is defined as the ratio of the number of BSs in the circular neighborhood to the area of the neighborhood. $f_{\rm{LoS}} \left( a,b,\lambda,r \right)$ is the PDF of the distance between the MRP and its associated BS,
\begin{equation}\label{PDF_of_dis}
\begin{split}
    & f_{\rm{LoS}} \left( a,b,\lambda,r \right) = 2\pi \lambda P_{\rm{LoS}}\left(a,b,r\right) \\
    & \times \exp\left( -2\pi \lambda \int_{h_{\rm{BS}}}^r l \, P_{\rm{LoS}}\left( a,b,l \right) \mathsf{d}l \right),
\end{split}
\end{equation}
$P_{\rm{LoS}}\left(a,b,r\right)$ is the LoS probability given in (\ref{PLoSk}), and $\mathcal{L} \left( a,b,\lambda,s \right)$ is Laplace transform of the interference plus noise,
\begin{sequation}\label{Laplace}
\begin{split}
    & \mathcal{L} \left( a,b,\lambda,s \right) = \exp \Bigg(- s\, \sigma B -2\pi \lambda \\
    & \times \int_r^{R_{\max}} \left[ 1 - \left( \frac{m_{\rm{LoS}}}{m_{\rm{LoS}} + s \, \eta_{\rm{LoS}} \rho \, l^{-\alpha_{\rm{LoS}}} } \right)^{m_{\rm{LoS}}}  \right] l \, P_{\rm{LoS}}\left(a,b,r\right) \mathsf{d}l 
    %-2\pi \lambda \int_{\max\left\{h, \left( \frac{\eta_{\rm{NLoS}}}{\eta_{\rm{LoS}}}\right)^{\frac{1}{\alpha_{\rm{NLoS}}}} l^{\frac{\alpha_{\rm{LoS}}}{\alpha_{\rm{NLoS}}}}  \right\}}^{R_{\max}} \left[ 1 - \left( \frac{m_{\rm{NLoS}}}{m_{\rm{NLoS}} + s \, \eta_{\rm{NLoS}} T \, l^{-\alpha_{\rm{NLoS}}} } \right)^{m_{\rm{NLoS}}}  \right] l \, \left( 1-P_{\rm{LoS}}\left(a,b,r\right) \right) \mathsf{d}l  
     \Bigg).
\end{split}
\end{sequation}

The features of the expression (\ref{SGcoverage}) can be summarized as follows:
\begin{itemize}
    \item The expression contains double integrals, and the computational complexity is high. Therefore, simplification is necessary.
    \item The coverage probability is considered as a function of $a, b$ and $\lambda$, because these three parameters all vary with location. 
    \item $a$ and $b$ occur together in the LoS probability $P_{\rm{LoS}}\left(a,b,r\right)$. That means we can try to separate $\lambda$ and $P_{\rm{LoS}}\left(a,b,r\right)$ to facilitate subsequent model simplification and training. 
\end{itemize}

\begin{proposition}\label{proposition1}
The expression in (\ref{SGcoverage}) can be simplified as follows,
\begin{equation}\label{simplifiedcov}
\begin{split}
    P_{SG}^C \left(a,b,\lambda \right) & = C_{{\rm{cov}},1}\left(a,b\right) \, \lambda \, \left( C_{{\rm{cov}},2}\left(a,b\right) \right)^{\lambda} \\
    & + C_{{\rm{cov}},3}\left(a,b\right) \, \lambda \, \left( C_{{\rm{cov}},4}\left(a,b\right) \right)^{\lambda},
\end{split}
\end{equation}
where $C_{{\rm{cov}},k}\left(a,b\right), k \in \{1,2,3,4\}$ are terrain-related coefficients.
\begin{IEEEproof}
See Appendix~\ref{appA}.
\end{IEEEproof}
\end{proposition}

\subsection{Model Training}
In the previous subsection, we realized the separation of the terrain-related parameters $a$ and $b$ from the BS density $\lambda$ in the coverage probability expression. However, it can be seen from Appendix~\ref{appA} that the relationships between $C_{{\rm{cov}},k}\left(a,b\right), k \in \{1,2,3,4\}$ and terrain-related parameters are not explicit. Therefore, we propose a model training algorithm to explain how to train the value of $C_{{\rm{cov}},k}\left(a,b\right), k \in \{1,2,3,4\}$ under different combinations of $a$ and $b$. Since the training is not real-time, complexity analysis is not required. Readability and conciseness are more important than the preciseness of the algorithm, so we tend to use descriptions rather than formulas.

\begin{algorithm}[!ht] 
	\caption{Model Training Algorithm}
	\label{Alg4}
	\begin{algorithmic} [1]
		\STATE Fit the function of $b$ with $a$ by the least squared method, according to the typical $a$ and $b$ pairs provided in \cite{series2013propagation}.
  
        \STATE Take a set of terrain-related parameter pairs on the fitting function, denoted as $(a_1,b_1), (a_2,b_2), \dots, (a_{n_{ab}},b_{n_{ab}})$.
        
        \STATE Take a set of BS densities $\lambda_1, \lambda_2, \dots, \lambda_{n_{\lambda}}$. 
        
        \FOR{$i = 1 : n_{ab}$}
        \STATE Set the initial values of $\widetilde{l}_1$ and $\widetilde{l}_2$ as $\underset{l}{\argmax} \ 2 \pi l P_{\rm{LoS}}\left(a_i,b_i,l\right)$. Set the initial values of $\widetilde{r}$ as $\underset{r}{\argmax} \ 2 \pi r f_{\rm{LoS}}\left(a_i,b_i,\overline{\lambda},r\right)$, where $\overline{\lambda}=\sum_{j=1}^{n_\lambda} \lambda_j$. $\widetilde{l}_1$, $\widetilde{l}_2$ and $\widetilde{r}$ are defined in (\ref{app:formu2}).
        
        \STATE Substitute $\widetilde{l}_1$, $\widetilde{l}_2$ and $\widetilde{r}$ into (\ref{app:formu2}) to get the initial values of $C_{{\rm{cov}},k}^i, k \in \{1,2,3,4\}$.
        
        \STATE Record $C_{{\rm{cov}},k}^i, k \in \{1,2,3,4\}$ that minimize $\sum_{j=1}^{n_{\lambda}} \left( P_{SG}^C (a_i,b_i,\lambda_j) - P_{\rm{Simu}}^C (a_i,b_i,\lambda_j) \right)^2$, where $P_{SG}^C (a_i,b_i,\lambda_j)$ is defined in (\ref{simplifiedcov}) and $P_{\rm{Simu}}^C (\lambda)$ is the accurate coverage probability under the symmetric scenario obtained through Monte Carlo simulation \cite{alzenad2019coverage}.
        
        \ENDFOR
        
		\STATE \textbf{Output}: The values of $C_{{\rm{cov}},k}^i, k \in \{1,2,3,4\}$ for all of the terrain-related parameter pairs.
	\end{algorithmic}
\end{algorithm}	

\par
\begin{figure}[t]
\centering
\includegraphics[width=0.6\linewidth]{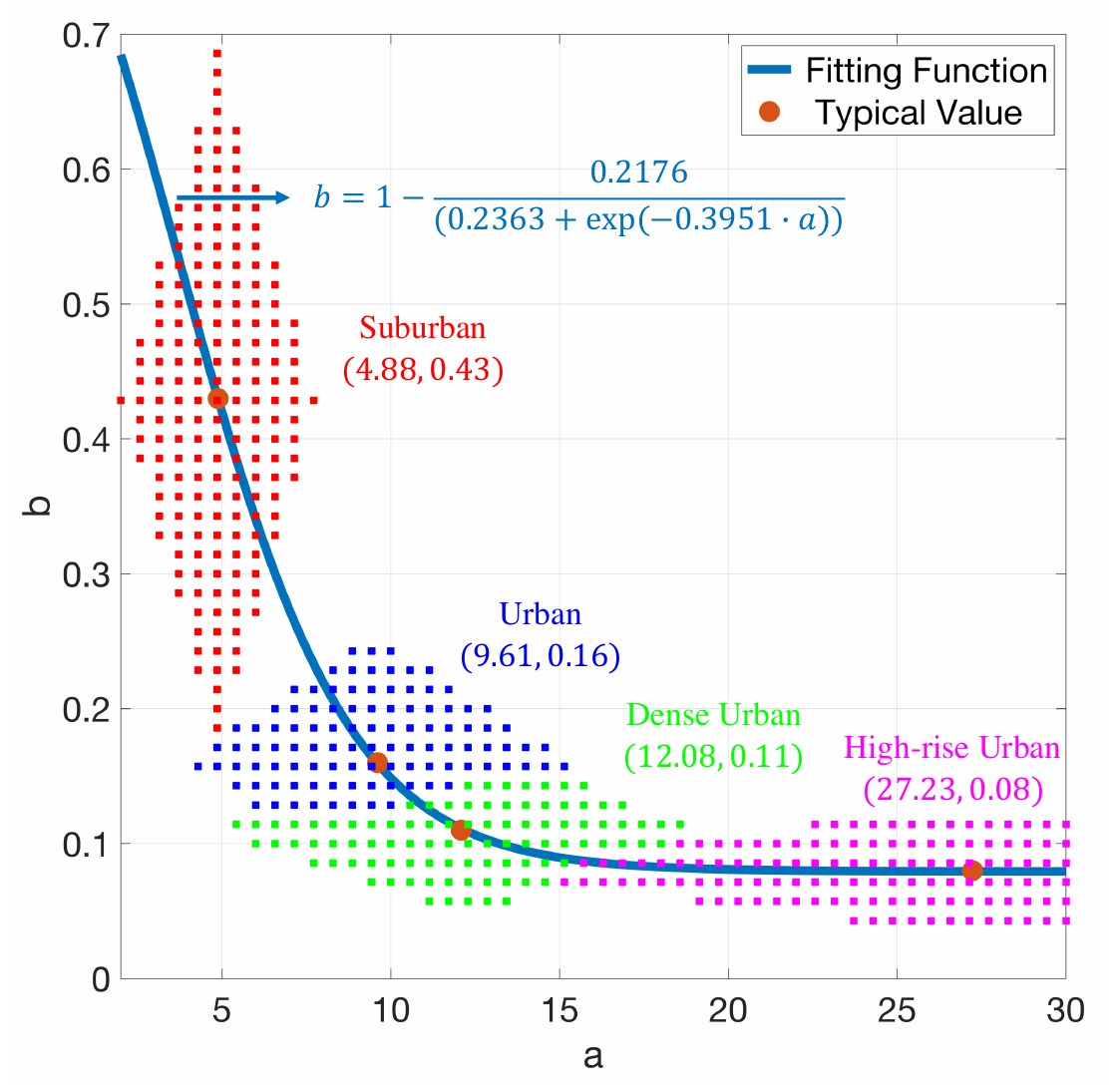}
\caption{Fitting function of $b$ with $a$ and application ranges of typical scenarios.}
\label{figure4}
\end{figure}

We choose a modified Sigmoid function to fit the function of $b$ with $a$ in algorithm~\ref{Alg4} step (1). The values of typical $a$ and $b$ pairs and the fitting function are given in Fig.~\ref{figure4}. In step (5), we set the distance corresponding to the strongest total interference power and the distance where the associated BS is most likely to locate as the initial values of $\widetilde{l_Q}, Q=\{1,2\}$ and $\widetilde{r}$. Finally, we apply the trust region reflective algorithm to minimize the well-known $\mathcal{L}^2$ norm loss function in step (7).

\begin{figure}[t]
\centering
\includegraphics[width=0.6\linewidth]{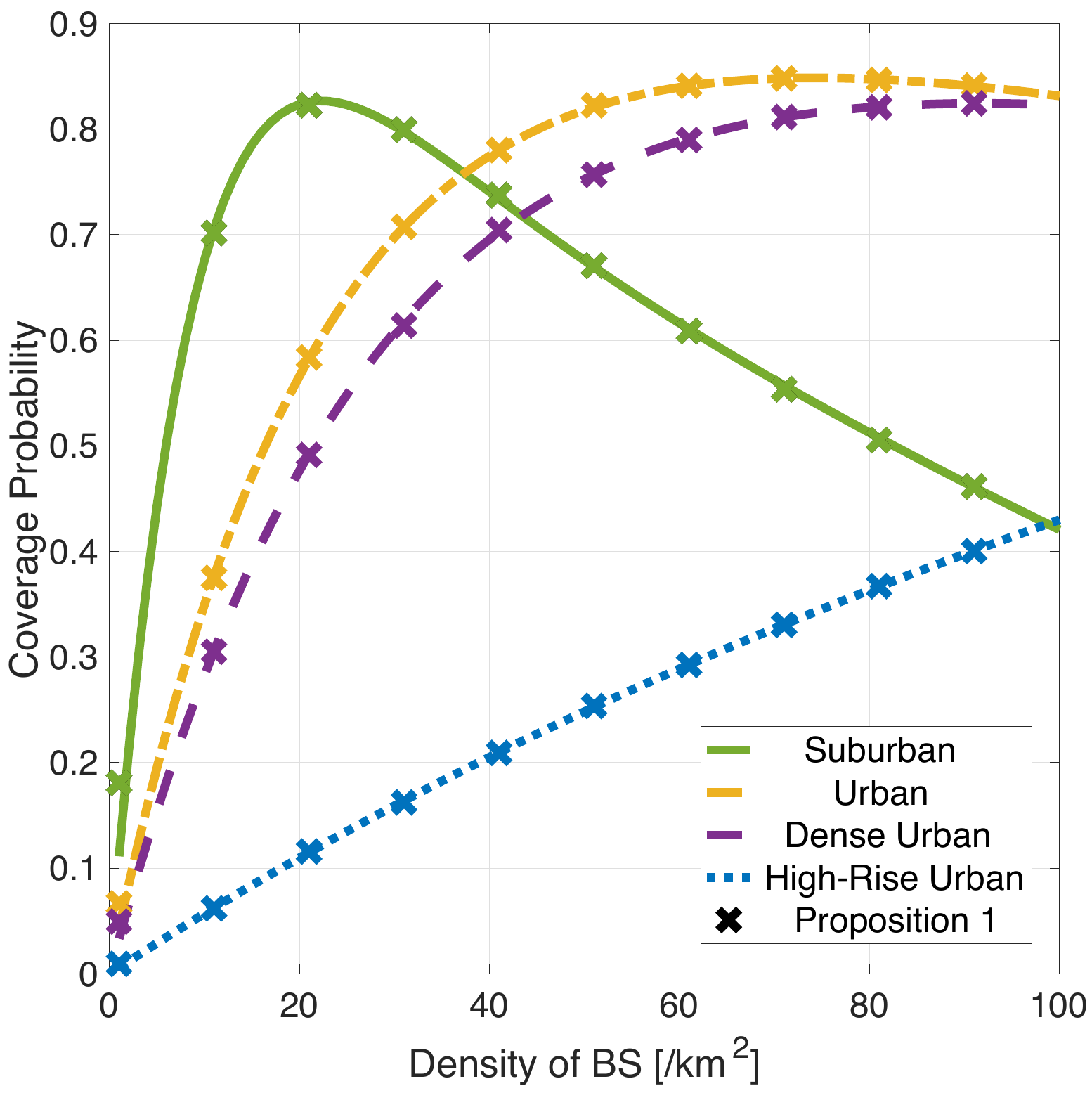}
\caption{Accuracy verification of the simplified SG-based expression.}
\label{figure5}
\end{figure}

We set up a database to record the trained terrain-related coefficients. The values of the coefficients under four typical scenarios defined in \cite{series2013propagation} are given in Table~\ref{table2}. The overlap of crosses and curves in Fig.~\ref{figure5} shows the accuracy of the simplified expression given in (\ref{simplifiedcov}) under these four scenarios. The curves are obtained by analytical expression proposed in \cite{alzenad2019coverage} under symmetric scenarios. The crosses in Fig.~\ref{figure5} are obtained by the simplified closed-form expression in (\ref{simplifiedcov}).

\begin{table}[]
\centering
\caption{The values of the trained terrain-related coefficients under typical scenarios.}
\label{table2}
\begin{tabular}{|c|c|c|c|c|}
\hline
      & Suburban & Urban   & Dense Urban & High-Rise Urban \\ \hline \hline
$a$   & 4.88     & 9.61    & 12.08       & 27.23           \\ \hline
$b$   & 0.43     & 0.16    & 0.11        & 0.08            \\ \hline
$C_{{\rm{cov}},1}$ & 0.0889   & 0.0271  & 0.0206      & 0.0028          \\ \hline
$C_{{\rm{cov}},2}$ & 0.9315   & 0.9680  & 0.9744      & 0.9999          \\ \hline
$C_{{\rm{cov}},3}$ & -0.0290  & -0.0167 & -0.0133     & -0.0031         \\ \hline
$C_{{\rm{cov}},4}$ & 0.9807   & 0.9917  & 0.9932      & 0.9929          \\ \hline
\end{tabular}
\end{table}

\subsection{SG-Based Method and Complexity Analysis}
This subsection explains how to obtain the coverage probability through the simplified analytical expression. In step (3) of algorithm~\ref{Alg5}, $a_i$ and $b_i$ are typical pairs of values recorded in the database and set by step (2) in algorithm~\ref{Alg4}. Because the values of $\widehat{a}$ and $\widehat{b}$ differ several orders of magnitude, normalized deviation in step (3) is used to measure the difference between $(\widehat{a},\widehat{b})$ pair and $(a_i,b_i),i\in \{1,2,\dots,n_{ab}\}$ pairs. $i^*$ is the index corresponding to the closest pair to pair $(\widehat{a},\widehat{b})$ in the database. Step (4)-(9) in algorithm~\ref{Alg5} calculate the density of BS $\lambda$ according to its definition.

\begin{algorithm}[!ht] 
	\caption{SG-Based Method}
	\label{Alg5}
	\begin{algorithmic} [1]
 
        \STATE Execute step (1) in algorithm~\ref{Alg3} (Replace $P_{SG}^C$ into $P_{\rm{AcSimu}}^C$). 
		
		\STATE Execute step (2)-(8) in algorithm~\ref{Alg3} to obtain $\widehat{a}$ and $\widehat{b}$ corresponding to the terrain. 

        \STATE $i^* \leftarrow \underset{i}{\argmin} \frac{\left| a_i - \widehat{a} \right|}{\widehat{a}} + \frac{\left| b_i - \widehat{b} \right|}{\widehat{b}}$.

        \STATE $\lambda \leftarrow 0$.
        \FOR{$i = 1 : M$}
        \IF{$\left(x_{{\rm{BS}},i} - x_{\rm{MRP}} \right)^2 + \left(y_{{\rm{BS}},i} - y_{\rm{MRP}} \right)^2 < R_\odot^2$}
        \STATE $\lambda \leftarrow \lambda + \frac{1}{\pi R_\odot^2}$.
        \ENDIF
        \ENDFOR
        
        \STATE $P_{SG}^C \leftarrow C_{{\rm{cov}},1}^{i^*} \ \lambda \ \left( C_{{\rm{cov}},2}^{i^*} \right)^{\lambda} + C_{{\rm{cov}},3}^{i^*} \ \lambda \ \left( C_{{\rm{cov}},4}^{i^*} \right)^{\lambda}$.
		
		\STATE \textbf{Output}:  $P_{SG}^C$.
	\end{algorithmic}
\end{algorithm}

Overall, the SG-based method exchanges storage space to decrease complexity through model training. As analyzed in Sec.~\ref{section3}, the complexities of steps (1) and (2) in algorithm~\ref{Alg5} are both $\mathcal{O}(M)$ on average. Step (3) has a complexity of $\mathcal{O}\left( n_{ab} \right)$, and there are $n_{ab}$ normalization deviations that need to be obtained. The complexity of steps (4)-(9) is $\mathcal{O}(N)$ since each BS is faced with a round of judgment in density calculation. Therefore, the computational complexity of the SG-based method is $\mathcal{O} \left( M+n_{ab}+N \right)$, and unit complexity only involves simple operations.

\section{ML-Based Method}
In this section, we propose a data-driven method to improve the accuracy of coverage probability estimation by absorbing experience from data. This method is an improved linear regression method under the ML framework.

\subsection{Data Sets and Grid Model}
In this subsection, we systematically describe the data sets of the three methods. Sec.~\ref{sec2-1} provides the data format of BS location and terrain information. For MRPs in the same map (such as KAUST), the data of BS and terrain is shared. The location of each MRP and the shared data can constitute a testing sample of accelerated simulation or SG-based method. In this paper, we assume the locations of MRPs form an equally spaced grid topology. Therefore, accelerated simulation and the SG-based method take the position topology of MRPs and shared data as input and output coverage manifold at MRPs. Although one map can contain a large number of samples, we select multiple maps to reflect the diversity of the data. These maps involve different terrain features, including suburban, urban, dense urban, high-rise urban, and atypical regions. 

\par
Unlike the other two methods, the ML-based method's data set has a different purpose and a special form. Firstly, the data set of accelerated simulation or SG-based method is for testing purposes only. Note that the model training procedure in SG-based methods does not require data sets for training. However, ML-based methods require many training samples in addition to testing data. Secondly, if BS and terrain are recorded as the data format in Sec.~\ref{sec2-1}, a simple operation such as calculating the distances between BSs and building centers would result in $MN$ rounds of distance calculations. Therefore, we introduce the grid model to significantly reduce the complexity. The ML-based method's training and testing samples are composed of a BS position matrix $B$ and a terrain matrix $H$.

\begin{figure*}[h]
\centering
\includegraphics[width=0.99\linewidth]{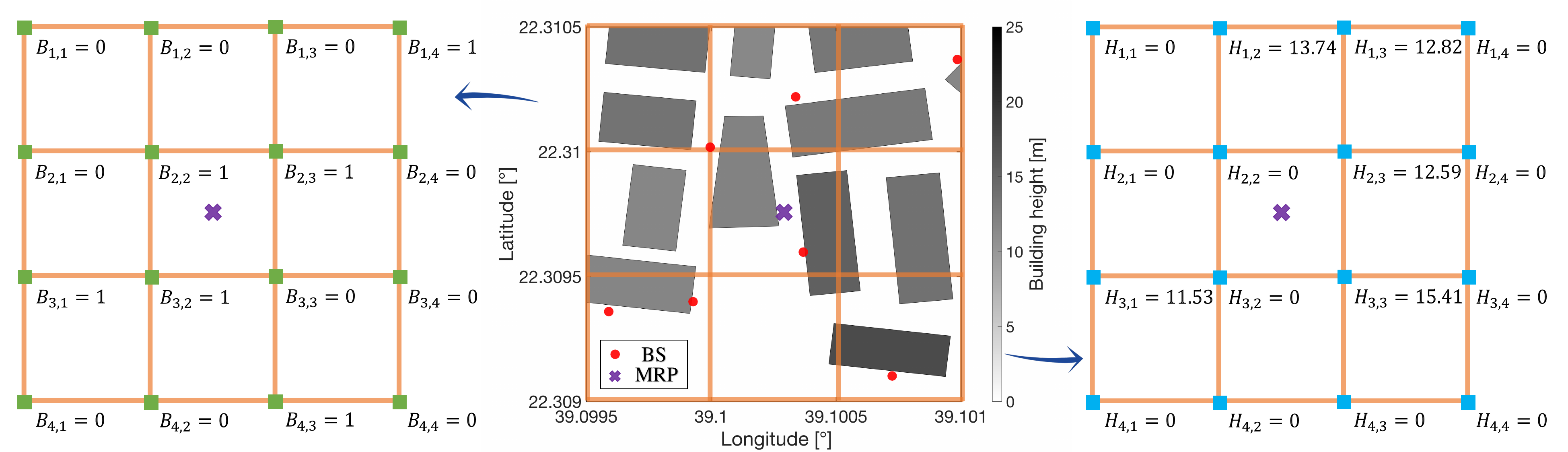}
\caption{An example of mapping from terrain to matrices.}
\label{figurematrix}
\vspace{-0.4cm}
\end{figure*}

\par
$B$ is a binary matrix whose element takes the value 1 if there is BS at the corresponding position and 0 otherwise. $B$ records the BS locations in a square neighborhood centered on MRP with side length $L$. $H$ records the height of the building corresponding to the same geographical location with $B$, thus $B$ and $H$ are of the same dimension. The dimension needs to be determined by balancing accuracy and complexity. The larger the dimension of $B$ and $H$, the higher the accuracy of the coverage probability, but the larger the computational complexity. To avoid the complex transformation from the system model established in Sec.~\ref{sec2-1} to the grid model, we prefer concise descriptions in the following algorithm. 

\begin{algorithm}[!ht] 
	\caption{The construction of Matrices $B$ and $H$}
	\label{Alg6}
	\begin{algorithmic} [1] 
        \STATE Perform type-I blockage verification for all positions from $H$ and construct $H$.
        
        \REPEAT
        \STATE Among the BSs that have not been removed, find the BS closest to the MRP. 
        \STATE Removed this closest BS.
        \STATE Perform type-II blockage verification for this BS. 
        \STATE If not blocked, this BS is the associated BS of MRP.
        \UNTIL{Find the associated BS of MRP.}
        
        \STATE Construct binary matrix $B$ to record the positions of BSs that have not been removed.
		
		\STATE \textbf{Output}: The constructed matrices $H$ and $B$.
	\end{algorithmic}
\end{algorithm}	

We take the example shown in Fig.~\ref{figurematrix}, where $B$ and $H$ are four-dimensional matrices. The BS at the lower right of the MRP is the closest one, and this BS-MRP link is not blocked. Therefore, step (3)-(8) in algorithm~\ref{Alg6} is only executed for one round. This BS is removed ($B_{1,3} \leftarrow 0$) and associated according to steps (5) and (7). In step (9), every BS that has not been removed is mapped to the nearest element in matrix $B$. However, in the case of a small dimension ($4 \times 4$), this construction procedure might lead to some errors, for instance, because the building closest to MRP is not represented in $H$. This eventually might lead to an inaccurate evaluation of the interference power. Therefore, to ensure accuracy, the size of matrices is $8 \times 8$ in implementation, and the region of MRP's neighborhood is also expanded.

\subsection{Data Training}
Recall that the SG-based method is a special linear regression method. If the influence of NLoS signal power on SINR is ignored, the $i^{th}$ interfering BS transmits the signal with the probability of $P_{\rm{LoS}}\left(a,b,d_i\right)$, while it does not transmit with the probability of $1-P_{\rm{LoS}}\left(a,b,d_i\right)$. The total average interference power under the A2GLPM $\sum_{i=1,i\neq i^*}^N P_{\rm{LoS}}\left(a,b,d_i\right) S_{{\rm{LoS}}, i}$ can be regarded as linear regression with the weight of $P_{\rm{LoS}}\left(a,b,d_i\right)$.

\par
Linear regression is suitable for the relationship modeling of the above scenario with few input features. Therefore, we can also design a linear regression method based on data training. Since the received power from each BS (independent variable) and total interference power (dependent variable) at the MRP are not difficult to obtain, we can train weights to replace the effect of the A2GLPM. Note that the coverage manifold of a region around MRP can be estimated with only two matrices, $B$ and $H$, as inputs. 

\par
The difference between SG and ML-based methods is: the SG framework determines weights by the statistical model, while the ML-based method trains the weights through the terrain matrix $H$. Same as the A2GLPM, the weight $w_{i,j}$ represents the probability that the BS at the position of $B_{i,j}$ establishes an LoS link with the MRP, which should satisfy the following three characteristics:
\begin{itemize}
    \item The value of $w$ is only affected by some elements in $H$ because not all buildings will block the BS-MRP link. In addition, the impact of the same building should only be considered once.
    \item Since each building blocks the link independently, $w$ can be expressed as a product of a set of LoS probabilities as functions of the building heights.
    \item The LoS probability through a single building should monotonically decrease as the height of the building increases. When the height of the building is 0, the LoS probability is 1. The curve of LoS probability is flat elsewhere but has a steep slope at a certain height.
\end{itemize}

We find that the modified Tanh function satisfies the third characteristic. Therefore, taking $w_{1,4}$ in Fig.~\ref{figurematrix} as an example,
\begin{equation}\label{w14}
\begin{split}
    w_{1,4} & = \left( \frac{1}{2} - \frac{1}{2}\tanh \left( \tau_{1,4}^{(1)} \left(H_{1,4}-h_{1,4}^{(1)}\right) \right) \right) \\
    & \times \left( \frac{1}{2} - \frac{1}{2}\tanh \left( \tau_{1,4}^{(2)} \left(H_{2,3}-h_{1,4}^{(2)}\right) \right) \right) \\
    & \times \left( \frac{1}{2} - \frac{1}{2}\tanh \left( \tau_{1,4}^{(3)} \left(H_{1,3}-h_{1,4}^{(3)}\right) \right) \right) \\
    & \times \left( \frac{1}{2} - \frac{1}{2}\tanh \left( \tau_{1,4}^{(3)} \left(H_{2,4}-h_{1,4}^{(3)}\right) \right) \right),
\end{split}
\end{equation}
where $\tau_{1,4}^{(1)}, \tau_{1,4}^{(2)}, \tau_{1,4}^{(3)}$ are scale parameters used to determine the descent rate of the modified Tanh function, and $h_{1,4}^{(1)}, h_{1,4}^{(2)}, h_{1,4}^{(3)}$ are the height thresholds that control when the Tanh functions decreases. The four terms in (\ref{w14}) represent the independent blockage of buildings or potential buildings in four positions according to $H_{1,4}, H_{2,3}, H_{1,3}, H_{2,4}$. Because of the symmetry, parameters $\tau_{1,4}^{(3)}$ and $h_{1,4}^{(3)}$ are equal for positions $H_{1,3}$ and $H_{2,4}$. In addition, if there is a duplicate blockage verification of the same building, one of the items in (\ref{w14}) must be removed. Taking $H_{2,3}=H_{2,4}$ as an example, we preferentially remove the influence of the fourth term in (\ref{w14}) because $H_{2,4}$ is further away from MRP compared to $H_{2,3}$. 

\par
Since many scale parameters and height thresholds need to be trained, a large number of accurate samples are required to support training. Therefore, accelerated simulation is used to get approximate training samples within an acceptable time. We will regenerate the building heights in the same map when necessary to expand the data set. Then, we make full use of symmetry to reduce the number of parameters. The proportion of unblocked samples of different maps at the same height is taken as the target value of learning. We obtain a series of target values by changing the height. As with model training, the trust region reflective algorithm is applied to minimize the $\mathcal{L}^2$ norm loss. Finally, there is a suggestion about the initial value of the height thresholds. The initial values of the height thresholds can be set according to the geometry relationship of the similar triangle in Fig.~\ref{figure3}. For the BS at the position of $B_{1,4}$ in Fig.~\ref{figurematrix}, the initial values can be set as follows: $H_{1,4}=h_{\rm{BS}}, H_{2,3}=\frac{1}{3} h_{\rm{BS}}$ and $H_{1,3}=H_{2,4}=\frac{2}{3} h_{\rm{BS}}$.

\subsection{ML-Based Method and Complexity Analysis}\label{section5-3}
In this subsection, we propose an ML-based method that refers to the core concept of the A2GLPM and makes relatively full use of terrain information. Then, the complexity of this method is analyzed. The following algorithm uses the trained weights to estimate the coverage probability at MRP. 

\begin{algorithm}[!ht] 
	\caption{ML-Based Method (For $4\times 4$ Matrices)}
	\label{Alg7}
	\begin{algorithmic} [1]
        \STATE \textbf{Input}: Matrices $B$ and $H$.
        \STATE Execute step (1) in algorithm~\ref{Alg3} (Replace $P_{ML}^C$ into $P_{\rm{AcSimu}}^C$). 

        \FOR{i = 1 : N}
        
        \STATE \textbf{Case 1} [The $i^{th}$ BS is associated]: $i^* \leftarrow i$, $\overline{S}_i \leftarrow \rho \eta_{\rm{LoS}} d_i^{\alpha_{\rm{LoS}}}$.
        \STATE \textbf{Case 2} [The $i^{th}$ BS is removed but not associated]: $\overline{S}_i \leftarrow \rho \eta_{\rm{NLoS}} d_i^{\alpha_{\rm{NLoS}}}$.
        \STATE \textbf{Case 3} [There is a mapping between the $i^{th}$ BS and a non-zero element in $B_{j,k} \neq 0$]: \\ $\overline{S}_i \leftarrow w_{j,k}(H) \rho \eta_{\rm{LoS}} d_i^{\alpha_{\rm{LoS}}} + \left(1-w_{j,k}(H)\right) \rho \eta_{\rm{NLoS}} d_i^{\alpha_{\rm{NLoS}}}$.
        \STATE \textbf{Else}: $\overline{S}_i \leftarrow 0$.
        \ENDFOR
        
        \STATE Execute step (18)-(25) in algorithm~\ref{Alg3} (Replace $P_{ML}^C$ into $P_{\rm{AcSimu}}^C$ in algorithm~\ref{Alg3} step (23)).
		\STATE \textbf{Output}: $P_{ML}^C$.
	\end{algorithmic}
\end{algorithm}	

In steps (4)-(7), we divide BSs into four categories: (\romannumeral1) The associated BS definitely establishes an LoS link with the MRP; (\romannumeral2) A not associated BS but closer than the associated one definitely establishes an NLoS link with the MRP; (\romannumeral3) For a BS that is farther than the associated one and located in the square neighborhood of MRP, the LoS BS-MRP link is established with the probability of $w_{i,j}(H)$; and (\romannumeral4) The received signal from a BS outside the square neighborhood is ignored.

\par
The computational complexities of step (2) and step (9) in algorithm~\ref{Alg7} are analyzed in Sec.~\ref{section3-3}. Note that for different MRP/samples of the same map, the height of the building at a specific geographic location is fixed. Therefore, the complexity of performing the type-I blockage verification can be shared by multiple MRPs in obtaining height matrix $H$. The complexity of obtaining the matrix $H$ can be ignored for a single MRP. As with the accelerated simulation method, the complexity analysis needs classified discussion:
\begin{itemize}
    \item In normal cases, the complexity of the simulation is determined by step (6) in algorithm~\ref{Alg7}. We take the obtaining of one weight $w$, which can be referred to (\ref{w14}), as unit complexity. The complexity of the ML-based method is $\mathcal{O}(\lambda L^2)$, where $\lambda$ is the density of BS and $L$ is the side length of MRP's square neighborhood. 
    \item One special case is when tall buildings are densely distributed (high-rise urban scenario), and the complexity mainly comes from the verification of associated and removed buildings in steps (4)-(5). Similar to the analysis in the normal case of accelerated simulation, we can get a complexity of $\mathcal{O}( \overline{\mathcal{A}} \, \iota \, \overline{d} \, \overline{r} \,\overline{n})$ with the verification of one side of the building as unit complexity, where
    \begin{equation}
        \overline{d} = \int_{h_{\rm{BS}}}^{R_{\max}} l \, f_{\rm{LoS}}(a,b,\lambda,l) \, \mathsf{d}l,
    \end{equation}
    \begin{sequation}
        \overline{\mathcal{A}} = \pi \lambda \int_{h_{\rm{BS}}}^{R_{\max}} \left(1 - P_{\rm{LoS}}\left( a,b,l \right) \right)l^2 \, f_{\rm{LoS}}(a,b,\lambda,l) \, \mathsf{d}l.
    \end{sequation}
    \item The same as accelerated simulation, another special case is when there are few buildings (suburban scenario) and high demand for coverage accuracy. The complexity of the ML-based method is determined by step (9) in algorithm~\ref{Alg7}, which is $\mathcal{O}(N\,n_{\rm{iter}})$ with unit complexity as the small-scale fading updating.
\end{itemize}

\section{Numerical Results}
\begin{figure}[t]
\centering
\includegraphics[width=0.6\linewidth]{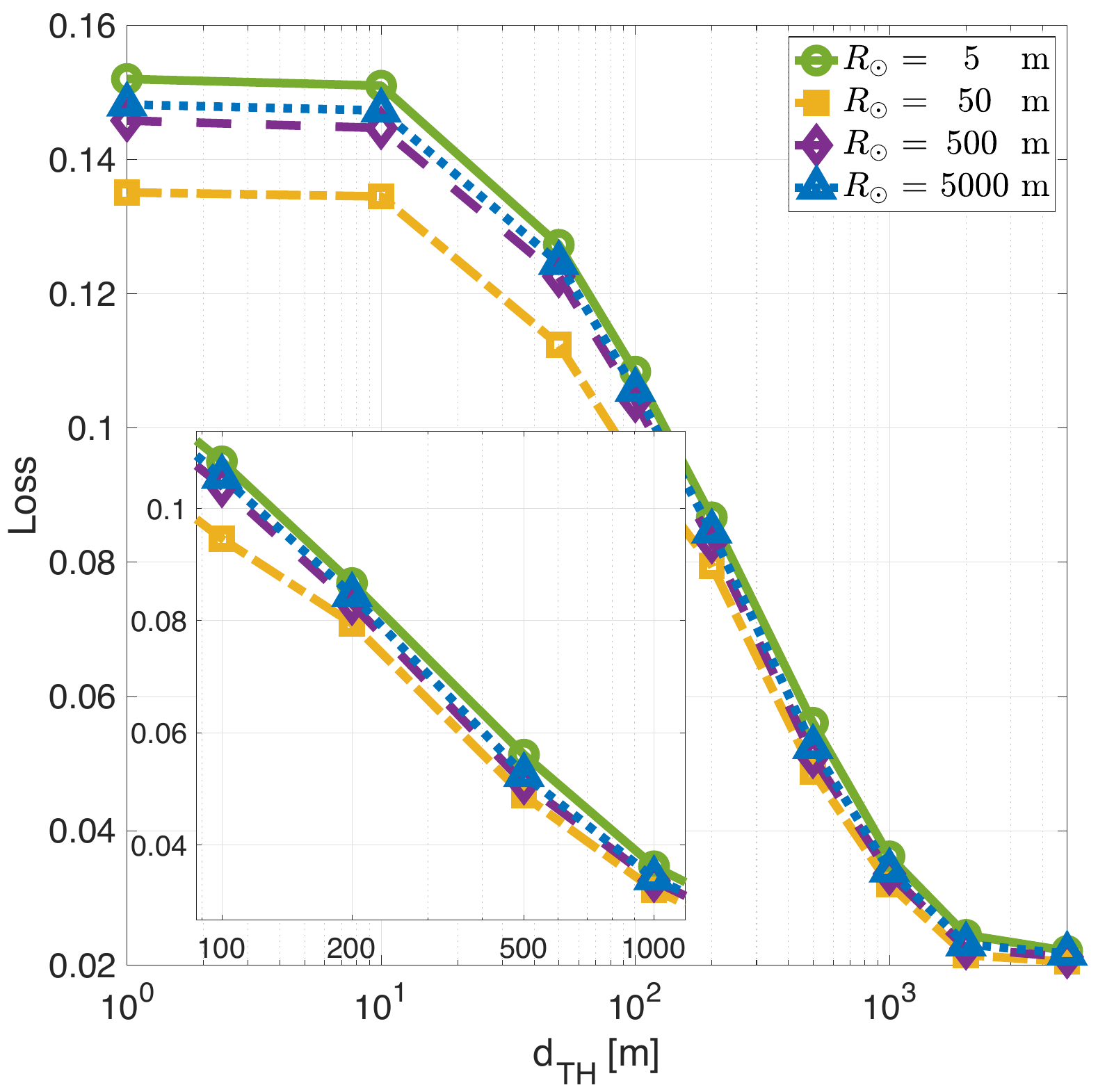}
\caption{Accuracy performance and parameter optimization of accelerated simulation.}
\label{figure7}
\end{figure}

In this section, we provide the numerical results accuracy and time delay of the three methods. In Fig.~\ref{figure7} and \ref{figure8}, we optimize the parameter in the accelerated simulation and SG-based method, respectively, and show their accuracy performance. In Fig.~\ref{figure9}-\ref{figure12}, the accuracy performances of the three methods under optimal parameters are compared. In Fig.~\ref{figure7}, \ref{figure8}, and \ref{figure9}, the loss is defined as the absolute value of the difference between the coverage probability at MRP obtained by the three methods and traditional simulation. Similarly, we excluded the testing samples with no BS or buildings in the MRP neighborhood to make the results more representative. Finally, we compare the time delay of the three methods in Fig.~\ref{figure13}.

\subsection{Accuracy Performances}
Recall that $d_{\rm{TH}}$ is the threshold that determines whether the A2GLPM is applied. With the increase of $d_{\rm{TH}}$, more BSs' blockages are strictly verified. The loss of coverage probability decreases linearly with the exponential increase of $d_{\rm{TH}}$ when $10\mathrm{m} < d_{\rm{TH}} < 1000 \mathrm{m}$. In the interval of $d_{\rm{TH}}>1000$m, it is no longer meaningful to further increase $d_{\rm{TH}}$. Next, when the A2GLPM is used to substitute the type-II blockage verification approximately, the model requires a pair of terrain-related parameters and BS density. $R_{\odot}$ determines the MRP's circular neighborhood of terrain-related parameters and BS density calculation. In Fig.~\ref{figure7}, when the horizontal radius of the circular neighborhood is selected as $R_{\odot}=5$m, the density of buildings and BSs may not be accurate since the area is too small, resulting in a loss in accuracy. When $R_{\odot}=5000$m, perhaps the entire map (for example, the map of KAUST in Fig.~\ref{figure1}) shares the same pair of parameters and BS density. Therefore, when $d_{\rm{TH}}$ is small, it is critical to select a moderate $R_{\odot}$. $R_{\odot}$ is set as $50$m in the following numerical results unless otherwise stated.

\begin{figure}[t]
\centering
\includegraphics[width=0.6\linewidth]{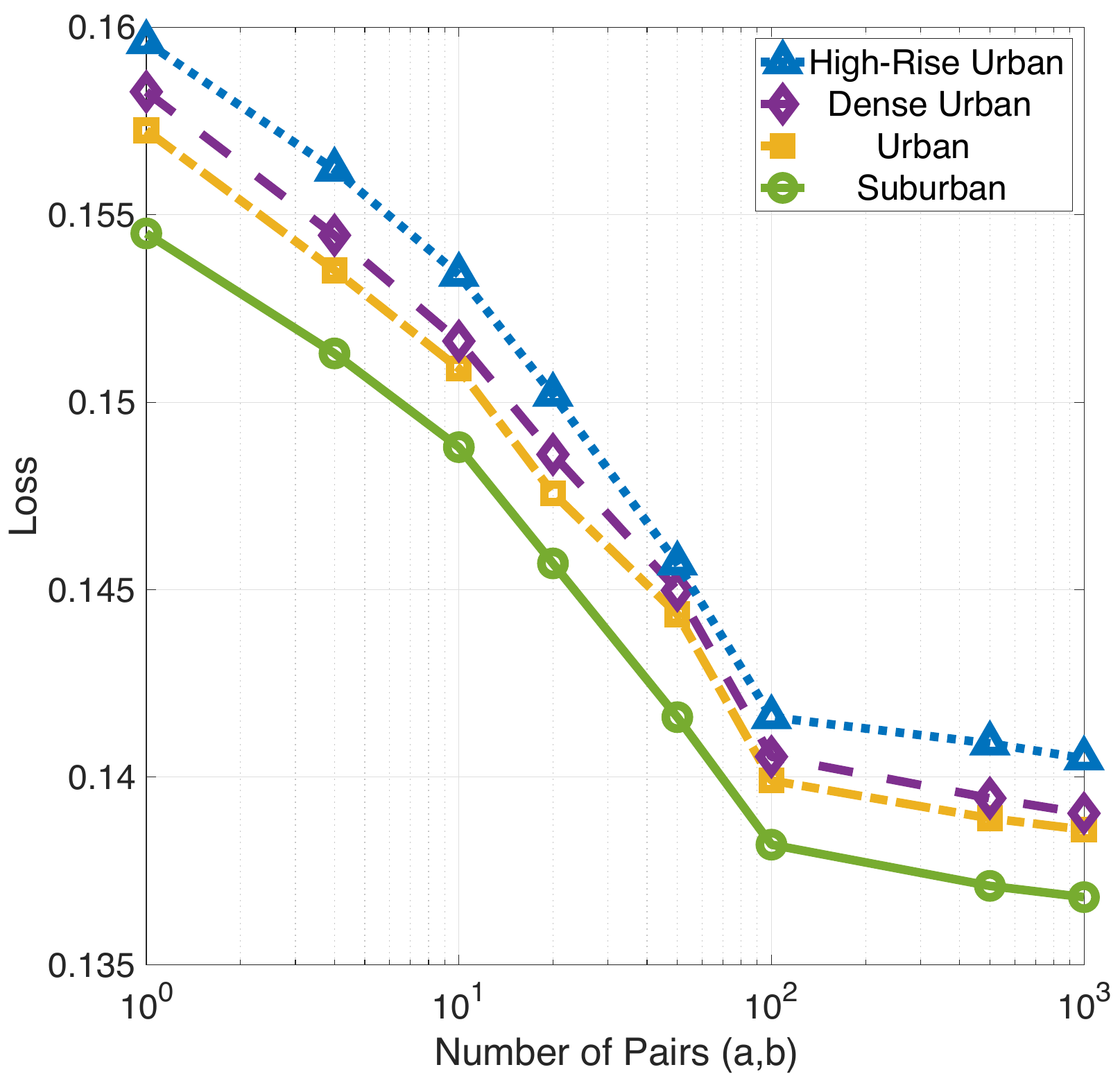}
\caption{Accuracy performance and parameter optimization of SG-based method.}
\label{figure8}
\end{figure}

\par
In Fig.~\ref{figure8}, overall, the loss of the SG-based method is significantly greater than that of accelerated simulation. Losses in four typical scenarios defined in Fig.~\ref{figure4} are respectively plotted in Fig.~\ref{figure8}. The coverage estimation of the SG-based method is more accurate in scenarios with smaller building densities and lower building heights. The meaning of the horizontal axis in Fig.~\ref{figure8} is the number of $(a,b)$ pairs obtained by model training and recorded in the database. For example, four $(a,b)$ pairs are recorded in Table~\ref{table2}. Performing elaborate model training and building a larger database is necessary when there are less than 100 terrain-related parameter pairs. With the further increase of the pair number, the loss gradually converges to the value of the accelerated simulation when $d_{\rm{TH}}=0$.

\begin{figure}[t]
\centering
\includegraphics[width=0.6\linewidth]{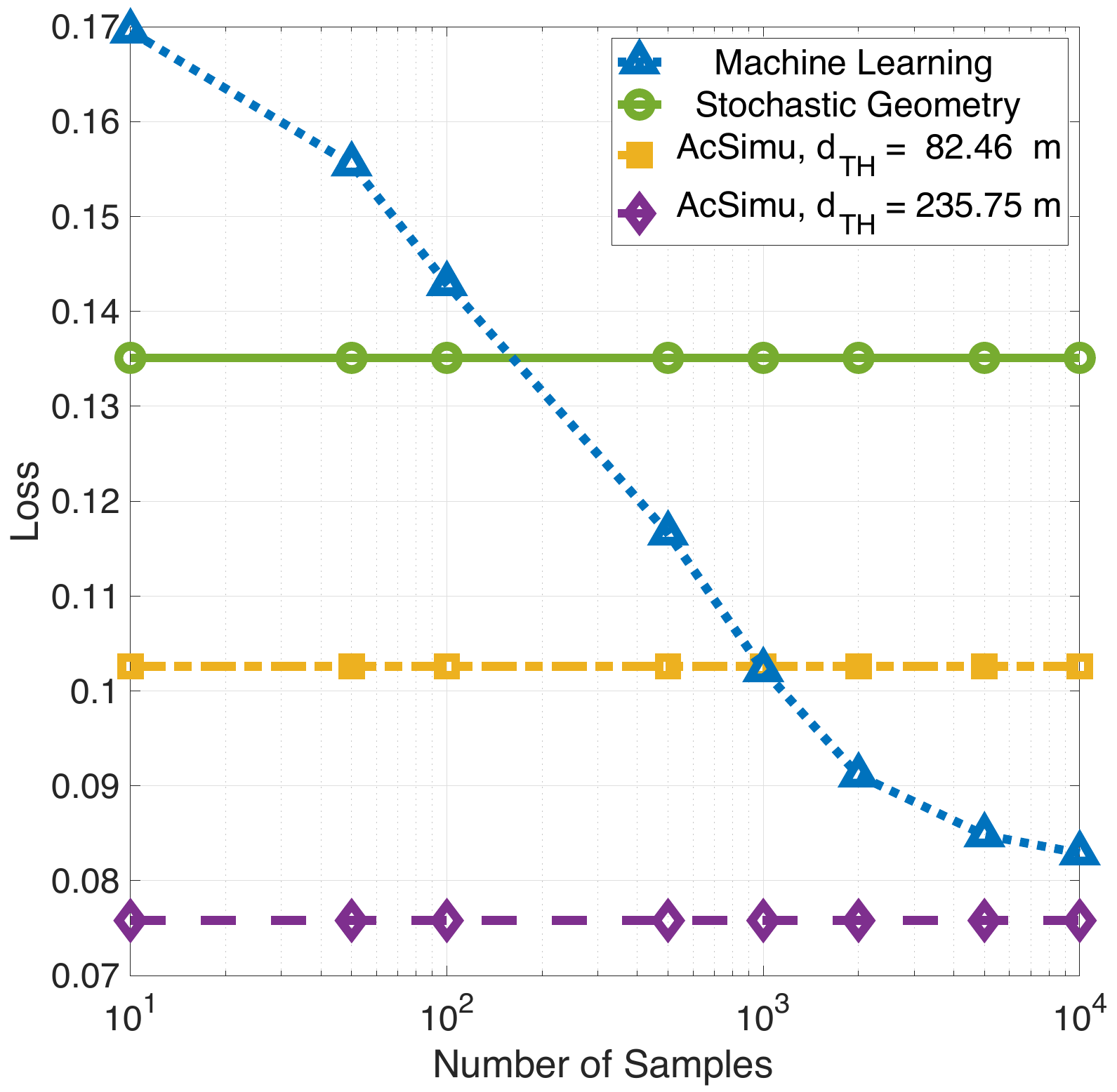}
\caption{Accuracy performance comparison of three methods.}
\label{figure9}
\end{figure}

\par
As shown in Fig.~\ref{figure9}, the loss of the ML-based method decreases rapidly as the number of training samples increases. Even compared with the loss of the SG-based method under the suburban scenario, the ML method only needs 200 training samples to obtain more accurate coverage probability estimation. In the legend of Fig.~\ref{figure9}, $82.46$m is the average distance between MRP and associated BS, while $235.75$m is the side length $L$ of the square neighborhood. For a BS in the circular neighborhood with radius $d_{\rm{TH}}=82.46$m, both accelerated simulation and ML-based method perform the type-II blockage verification. When more than 1000 samples are studied, the blockage experience learned from the data set is more accurate than that of the A2GLPM, and the loss obtained by the ML-based method is smaller than that obtained by accelerated simulation. The loss corresponding to the accelerated simulation with $d_{\rm{TH}}=235.75$m is a lower bound of the loss of the ML-based method because the ML-based method does not learn the terrain and BS location information outside the circular region with radius $d_{\rm{TH}}=235.75$m.

\begin{figure}[t]
\centering
\includegraphics[width=0.75\linewidth]{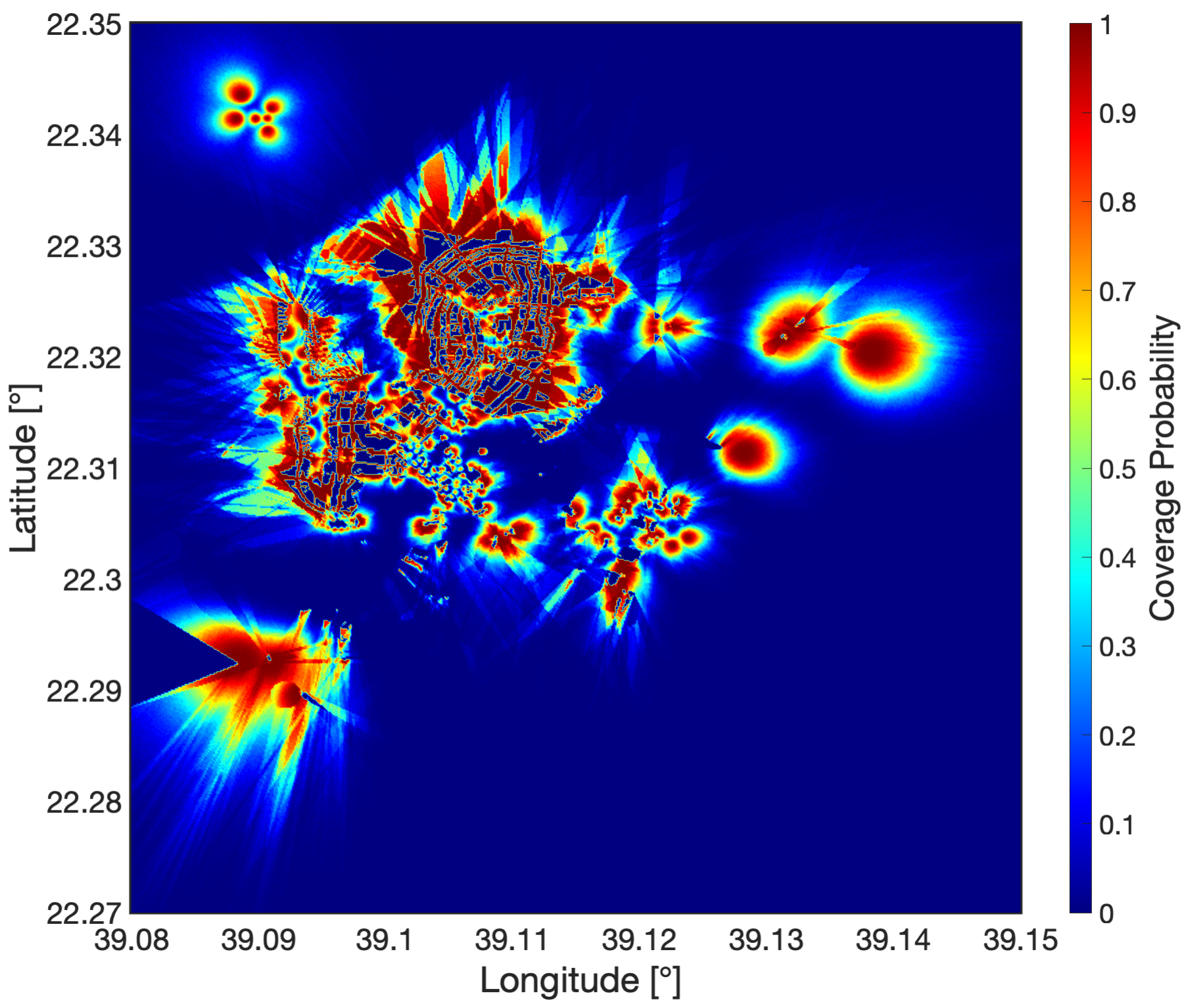}
\caption{Coverage manifold of KAUST (accelerated simulation).}
\label{figure10}
\end{figure}

\begin{figure}[t]
\centering
\includegraphics[width=0.75\linewidth]{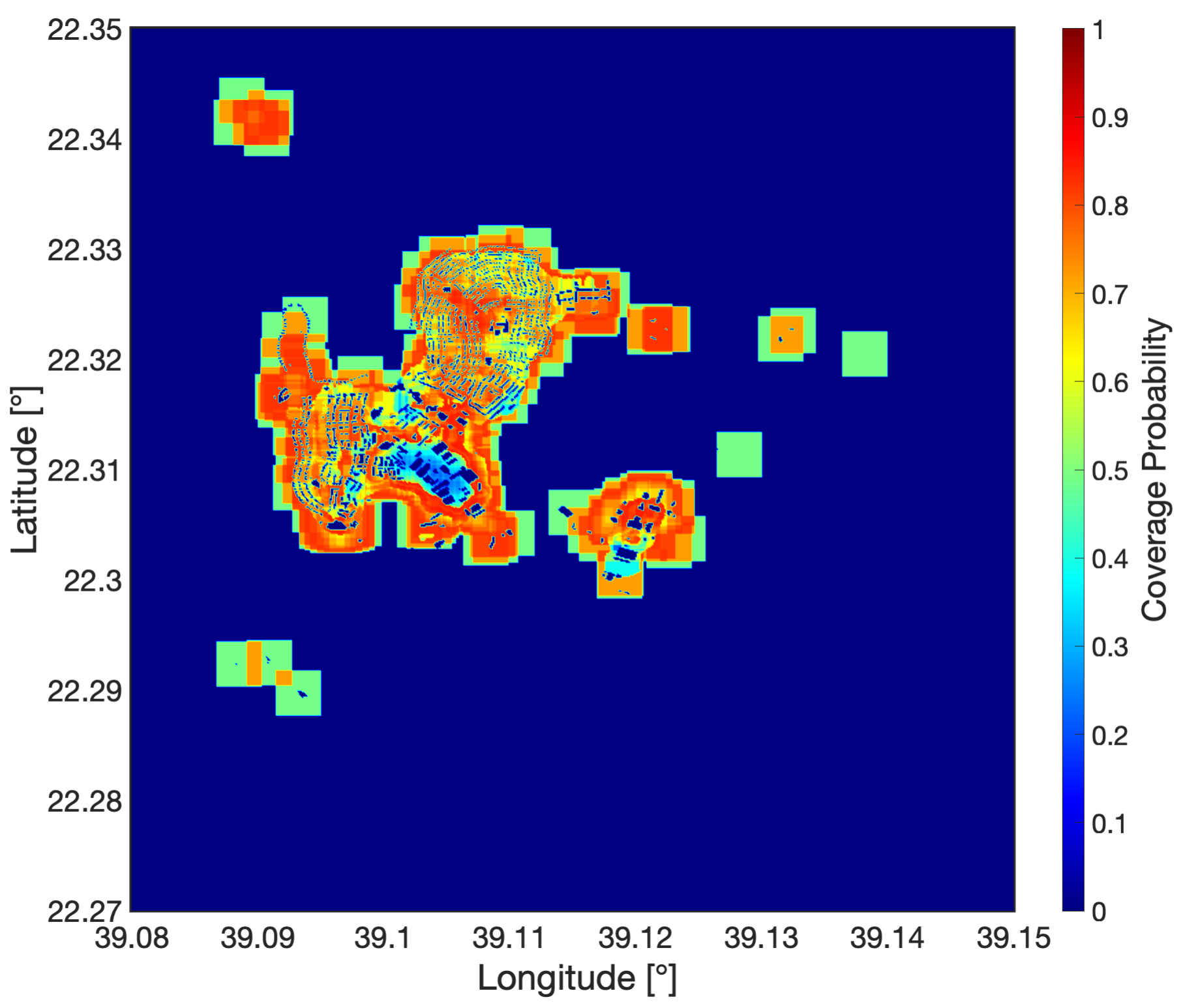}
\caption{Coverage manifold of KAUST (SG).}
\label{figure11}
\end{figure}

\subsection{Results of Coverage Manifold}
\begin{figure}[t]
\centering
\includegraphics[width=0.75\linewidth]{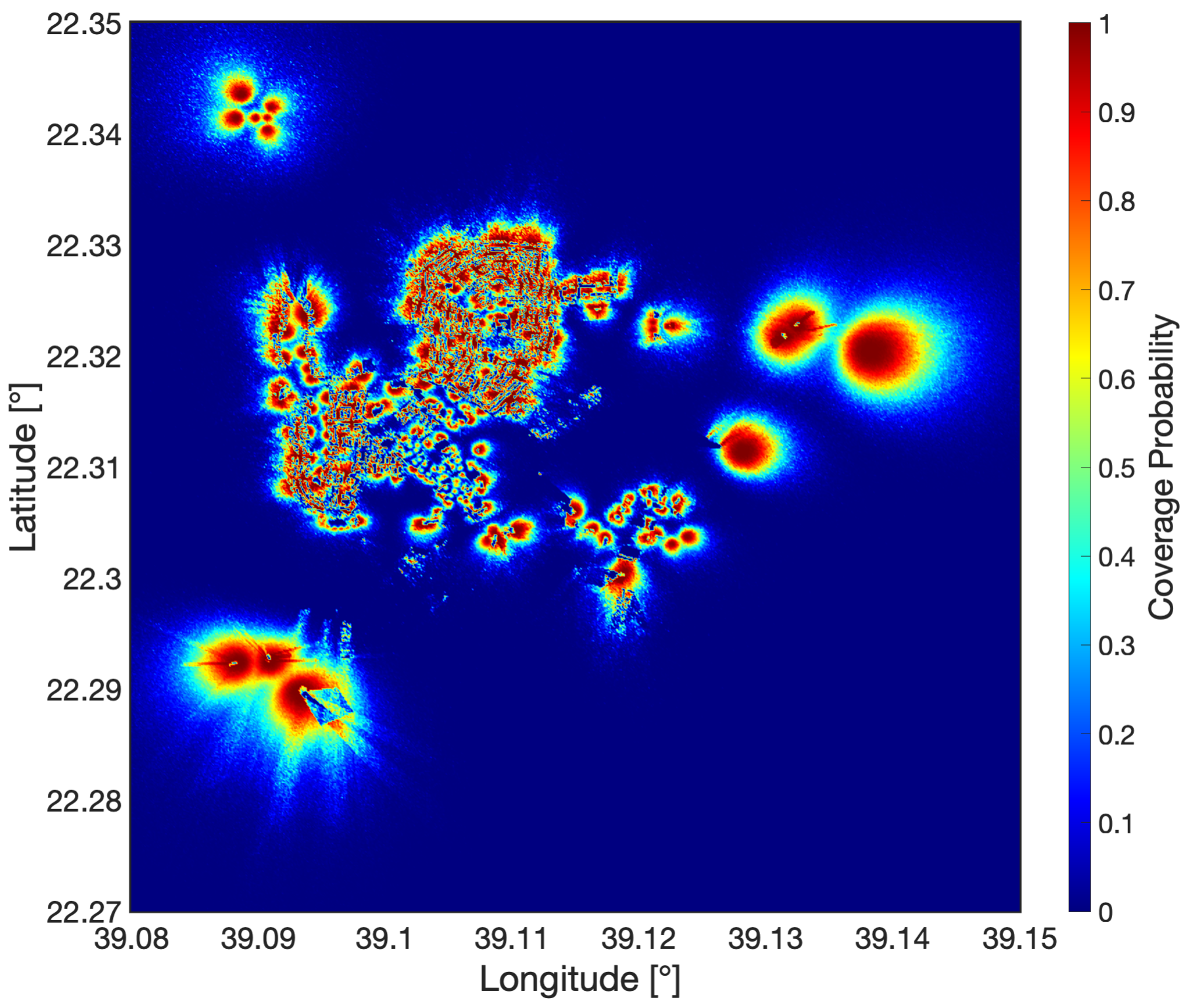}
\caption{Coverage manifold of KAUST (ML).}
\label{figure12}
\end{figure}

Fig.~\ref{figure10}, \ref{figure11}, and \ref{figure12} are coverage manifolds of KAUST obtained by accelerated
simulation, SG-based method, and ML-based method. $d-{\rm{TH}}$ is set as $1000$m in Fig.~\ref{figure10}, therefore the manifold is well estimated. In Fig.~\ref{figure11}, since the values of terrain-related parameter pairs and of BS densities of adjacent MRPs are similar, thus the coverage probability estimated by SG-based has a slight difference. This is the main reason the SG-based method is coarse-grained. In Fig.~\ref{figure12}, the manifold obtained by the ML-based method is similar to the manifold in Fig.~\ref{figure10}. The ML-based method learns the terrain and BS information in a limited neighborhood, thus the manifold has more 'burrs' and fewer 'corners' on the outer margin. They are respectively caused by the lack of associated BS in the neighborhood and wrong verification of blockage.

\begin{figure}[t]
\centering
\includegraphics[width=0.6\linewidth]{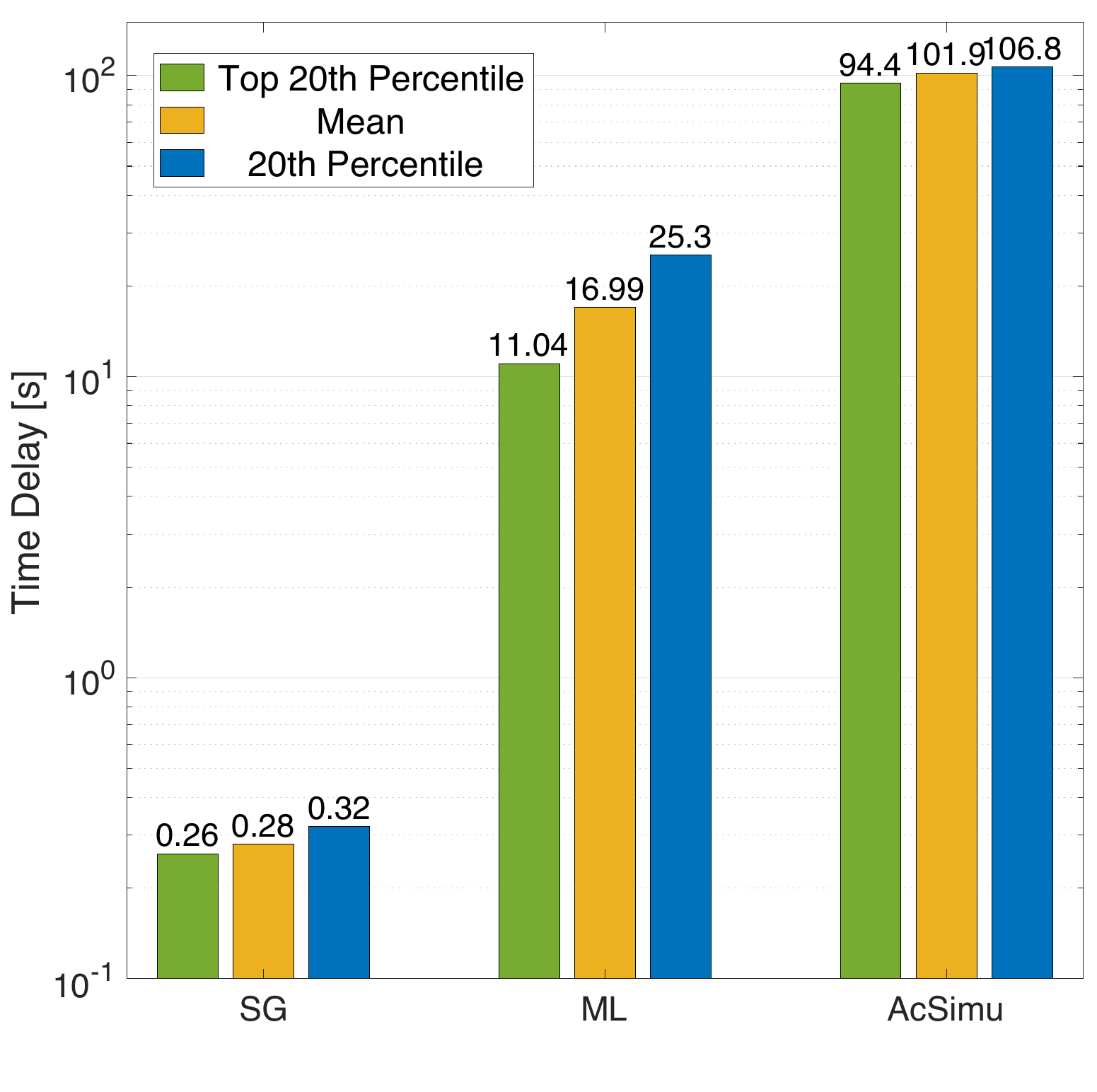}
\caption{Time delay performance comparison of three methods.}
\label{figure13}
\end{figure}

\subsection{Time Delay}
Fig.~\ref{figure13} compares the time delay required for a single MRP among the three methods. In the legend, the top 20th percentage and 20th percentage represent the average time delay of the samples with the shortest and longest time delay, respectively. The three methods have significant differences in time delay, among which the SG-based method has an obvious advantage in complexity compared with the other two methods. In addition, as previously analyzed, the delay of the SG-based method is almost unaffected by terrain and BS density. The ML-based method, on the other hand, has large fluctuations in time delay performance, which can be explained by the classified discussion in Sec.~\ref{section5-3}. Finally, through testing with nearly a hundred samples, the traditional simulation takes one to one and a half hours to complete the coverage manifold computation for a single MRP.

\section{Conclusion}
The traditional simulation is computationally expensive when faced with the task of coverage manifold estimation based on terrain. Therefore, we propose three methods as substitutions. The SG-based method trades storage space for complexity through model training. It has a minimal computational delay but is coarse-grained due to the compression of terrain and BS information. During the setup of a temporary network, this method can quickly provide parameters for network configuration, such as BS density. The ML-based method can balance the complexity and accuracy and is a compromise between accelerated simulation and the SG-based method. It is more suitable for the preliminary design of BS redeployment plans when there are significant changes in terrain topology or network structure. 
Moreover, for researchers with limited computing power, using the ML-based method to obtain manifold coverage can be a viable choice. The accelerated simulation can significantly reduce the complexity without significant loss of accuracy. However, accelerated simulation is still time-consuming because it cannot inherit the previous experience (such as model, data, etc.). Considering that future networks may involve a large number of signal sources and face more complex deployment environments, as well as possess network self-evolving capabilities \cite{letaief2019roadmap}, the computational complexity benefits of this method may still be meaningful even for network operators equipped with powerful computing capabilities.

\par
In future work, there are two potential directions to expand upon the research in this article. Firstly, we could extend the analysis of planar coverage manifold to spherical surfaces. This would involve modifying simulation blockage verification, designing spherical A2GLPM, and remodeling the distribution of buildings and BSs. Secondly, the three methods presented in this article were designed for scenarios with dense distributions of buildings and BSs. However, in remote areas, blockage might no longer be the primary limiting factor for performance. Thus, it would be plausible to combine the proposed methods with those outlined in \cite{mondal2022deep} to establish a system capable of providing coverage manifold in any scenario.

\appendices
\section{Proof of Proposition~\ref{proposition1}}\label{appA}
For convenience, the scale parameter of Nakagami-m fading is substituted as $m_{\rm{LoS}}=2$ into (\ref{SGcoverage}), and we get
\begin{sequation}\label{app:formu1}
\begin{split}
    & P^C \left(a,b,\lambda \right) \\
    & = \underbrace{ 2 \int_{h_{\rm{BS}}}^{R_{\max}} f_{\rm{LoS}} \left( a,b,\lambda,r \right) \mathcal{L} \left( a,b,\lambda, \sqrt{2} \gamma \, \rho^{-1} \, \eta_{\rm{LoS}}^{-1} r^{\alpha_{\rm{LoS}}} \right) \mathsf{d} r}_{{\rm{First \, term: \, }} P_1^C \left(a,b,\lambda \right)} \\
    & \underbrace{ - \int_{h_{\rm{BS}}}^{R_{\max}} f_{\rm{LoS}} \left( a,b,\lambda,r \right) \mathcal{L} \left( a,b,\lambda, 2\sqrt{2} \gamma \, \rho^{-1} \, \eta_{\rm{LoS}}^{-1} r^{\alpha_{\rm{LoS}}} \right) \mathsf{d} r}_{{\rm{Second \, term: \, }} P_2^C \left(a,b,\lambda \right)},
\end{split}
\end{sequation}
Substituting (\ref{PDF_of_dis}) and (\ref{Laplace}) into (\ref{SGcoverage}) and applying simple transposition, the first term of (\ref{app:formu1}) is given in (\ref{app:formu2}) at the top of next page,
\begin{table*}
\begin{equation}\label{app:formu2}
\begin{split}
    & P_1^C \left(a,b,\lambda \right) = 2 \int_{h_{\rm{BS}}}^{R_{\max}} 2\pi \lambda P_{\rm{LoS}}\left(a,b,r\right) \exp\left( -2\pi \lambda \int_{h_{\rm{BS}}}^r l \, P_{\rm{LoS}}\left( a,b,l \right) \mathsf{d}l \right) \\
    & \times \exp \Bigg(-2\pi \lambda \int_r^{R_{\max}} \left[ 1 - \left( \frac{m_{\rm{LoS}}}{m_{\rm{LoS}} + \sqrt{2} \gamma \,
    r^{\alpha_{\rm{LoS}}}  \, l^{-\alpha_{\rm{LoS}}} } \right)^{m_{\rm{LoS}}}  \right] l \, P_{\rm{LoS}}\left(a,b,r\right) \mathsf{d}l - s\, \sigma B \Bigg) \mathsf{d} r \\
    & \overset{(a)}{=} \underbrace{4 \pi \left( {R_{\max}} - {h_{\rm{BS}}} \right) P_{\rm{LoS}}\left(a,b,\widetilde{r}\right) \exp\left( -\sqrt{2} \gamma \, \rho^{-1} \, \eta_{\rm{LoS}}^{-1} \, \widetilde{r}^{\alpha_{\rm{LoS}}} \, \sigma B \right)}_{C_{{\rm{cov}},1}\left(a,b\right)}\times \lambda \\
    &\times \underbrace{\exp\left( -2\pi \left(\widetilde{r} - h_{\rm{BS}}\right) \widetilde{l_1} \, P_{\rm{LoS}}\left( a,b,\widetilde{l_1} \, \right) \right)^{\lambda}}_{\left(C_{{\rm{cov}},2}^{(1)}\left(a,b\right)\right)^{\lambda}} \\
    &\times \underbrace{ \exp \Bigg(-2\pi \left({R_{\max}} - \widetilde{r}\right) \left[ 1 - \left( \frac{m_{\rm{LoS}}}{m_{\rm{LoS}} + \sqrt{2} \gamma \,
    \widetilde{r}^{\alpha_{\rm{LoS}}}  \, \widetilde{l_2}^{-\alpha_{\rm{LoS}}} } \right)^{m_{\rm{LoS}}}  \right] \widetilde{l_2} \, P_{\rm{LoS}}\left(a,b,\widetilde{r}\right) \Bigg)^{\lambda}}_{\left(C_{{\rm{cov}},2}^{(2)}\left(a,b\right)\right)^{\lambda}},
\end{split}    
\end{equation}
\hrule
\end{table*}

\begin{table*}
\begin{equation}\label{app:formu3}
\begin{split}
    \mathcal{L} \left( a,b,\lambda,s \right) & = \exp \Bigg(-2\pi \lambda \int_r^{R_{\max}} \left[ 1 - \left( \frac{m_{\rm{LoS}}}{m_{\rm{LoS}} + s \, \eta_{\rm{LoS}} \rho \, l^{-\alpha_{\rm{LoS}}} } \right)^{m_{\rm{LoS}}}  \right] l \, P_{\rm{LoS}}\left(a,b,r\right) \mathsf{d}l \\
    & -2\pi \lambda \int_{\max\left\{h, \left( \frac{\eta_{\rm{NLoS}}}{\eta_{\rm{LoS}}}\right)^{\frac{1}{\alpha_{\rm{NLoS}}}} l^{\frac{\alpha_{\rm{LoS}}}{\alpha_{\rm{NLoS}}}}  \right\}}^{R_{\max}} \left[ 1 - \left( \frac{m_{\rm{NLoS}}}{m_{\rm{NLoS}} + s \, \eta_{\rm{NLoS}} \rho \, l^{-\alpha_{\rm{NLoS}}} } \right)^{m_{\rm{NLoS}}}  \right] \\
    & \times l \, \left( 1-P_{\rm{LoS}}\left(a,b,r\right) \right) \mathsf{d}l  
    - s\, \sigma B \Bigg),
\end{split}
\end{equation}
\hrule
\end{table*}

where $(a)$ follows the mean value theorem for double integrals \cite{young1917multiple}. Since $f_{\rm{LoS}} \left( a,b,\lambda,r \right)$ and $\mathcal{L} \left( a,b,\lambda,s \right)$ are continuous functions with respect to any integrated variable, it is easy to know that $P^C \left(a,b,\lambda \right)$ changes continuously when the integrated variable changes continuously. Therefore, there exists at least one set of $\left(\widetilde{r},\widetilde{l_1},\widetilde{l_2} \right)$ that makes the equation satisfied. Denote $C_{{\rm{cov}},2}\left(a,b\right)=C_{{\rm{cov}},2}^{(1)}\left(a,b\right) C_{{\rm{cov}},2}^{(2)}\left(a,b\right)$, the first term of (\ref{app:formu1}) is derived. The derivation of the second term is similar to that of the first term, therefore omitted here. 
\par
Finally, we give two supplements to the simplified coverage probability expression. As mentioned, when high-rise buildings are densely distributed, the power of NLoS BSs is not negligible. The Laplace transform of the interference in (\ref{Laplace}) is replaced by (\ref{app:formu3}) at the top of this page, and solving procedure of terrain-related coefficients $C_{{\rm{cov}},k}\left(a,b\right), k \in \{1,2,3,4\}$ follows the same steps. 
\par
Since the initial values of terrain-related coefficients $C_{{\rm{cov}},k}\left(a,b\right), k \in \{1,2,3,4\}$ are sensitive to the value of $R_{\max}$, the second supplement is about the value range of $R_{\max}$. When the distance between BS and MRP exceeds $R_{\max}$:
\begin{itemize}
    \item (i) BS has a high probability of being blocked, $P_{\rm{LoS}} \left(a,b,R_{\max}\right) < \epsilon_1$;
    \item (ii) The average NLoS received power is much lower than noise power, $\rho \eta_{\rm{NLoS}} R_{\max}^{\alpha_{\rm{NLoS}}} < \epsilon_2 \sigma B$;
    \item (iii) The probability that there is no available LoS BS is very low, $f_{\rm{LoS}}\left(a,b,\lambda,R_{\max} \right)>1-\epsilon_3$, where $f_{\rm{LoS}}\left(a,b,\lambda,R_{\max} \right)$ is the cumulative distribution function (CDF) of the distance between the MRP and its associated BS.
\end{itemize}

% \begin{table}[H]
% \centering
% \caption{Polynomial Coefficients for $b$}
% \label{app:table2}
% \begin{tabular}{|c|ccccc|}
% \hline
% $C_{ij}$ & $i$ & 0                    & 1                    & 2                    & 3                   \\ \hline
% $j$      &     &                      &                      &                      &                     \\
% 0        &     & $1.17\cdot 10^{0}$  & $-7.56\cdot 10^{-2}$  & $1.98\cdot 10^{-3}$ & $-1.78\cdot 10^{-5}$ \\
% 1        &     & $-5.79\cdot 10^{-3}$  & $1.81\cdot 10^{-4}$  & $-1.65\cdot 10^{-6}$  & $-$                 \\
% 2        &     & $1.73\cdot 10^{-5}$ & $-2.02\cdot 10^{-7}$ & $-$                  & $-$                 \\
% 3        &     & $-2.00\cdot 10^{-8}$  & $-$                  & $-$                  & $-$                 \\ \hline
% \end{tabular}
% \end{table}

\bibliographystyle{IEEEtran}
\bibliography{references}

\end{document}